\documentclass{aastex63}
\usepackage{graphicx}
\usepackage{CJK}
\received{----}
\revised{----}
\accepted{October 15, 2021}
\submitjournal{ApJ}

\begin{document}
\begin{CJK*}{UTF8}{gbsn}

\defcitealias{2021ApJ...907...18R}{Paper I}

\title{The Sample of Red Supergiants in Twelve Low-Mass Galaxies of the Local Group}

\correspondingauthor{Biwei Jiang}
\email{bjiang@bnu.edu.cn}

\author[0000-0003-1218-8699]{Yi Ren (任逸)}
\affiliation{Department of Astronomy, Beijing Normal University, Beijing 100875, People's Republic of China}
\affiliation{College of Physics and Electronic Engineering, Qilu Normal University, Jinan 250200, People's Republic of China}

\author[0000-0003-3168-2617]{Biwei Jiang (姜碧沩)}
\affiliation{Department of Astronomy, Beijing Normal University, Beijing 100875, People's Republic of China}

\author[0000-0001-8247-4936]{Ming Yang (杨明)}
\affiliation{IAASARS, National Observatory of Athens, Vas. Pavlou and I. Metaxa, Penteli 15236, Greece}

\author[0000-0001-5197-4858]{Tianding Wang (王天丁)}
\affiliation{Department of Astronomy, Beijing Normal University, Beijing 100875, People's Republic of China}

\author[0000-0003-4988-3513]{Tongtian Ren (任桐田)}
\affiliation{Department of Astronomy, Beijing Normal University, Beijing 100875, People's Republic of China}



\begin{abstract}
This work establishes the most complete sample of red supergiants (RSGs) in twelve low-mass galaxies (WLM, IC 10, NGC 147, NGC 185, IC 1613, Leo A, Sextans B, Sextans A, NGC 6822, Pegasus Dwarf, SMC and LMC) of the Local Group, which forms the solid basis to study the properties of RSGs as well as the star formation rate (SFR) and initial mass function (IMF) of the galaxies. After removing the foreground dwarf stars by their obvious branch in the near-infrared color-color diagram ($(J-H)_0/(H-K)_0$) with the UKIRT/WFCAM and 2MASS photometry as well as the Gaia/EDR3 measurements of proper motion and parallax, RSGs are identified from their location in the color-magnitude diagram $(J-K)_{0}/K_{0}$ of the member stars of the specific galaxy. A total of 2,190 RSGs are found in ten dwarf galaxies, and additionally 4,823 and 2,138 RSGs in the LMC and SMC respectively. The locations of the tip of the red giant branch in the $(J-K)_{0}/K_{0}$ diagram are determined to serve as an indicator of the metallicity and distance modulus of the galaxies.
\end{abstract}



\section{Introduction} \label{sec:intro}

Red supergiants (RSGs) are Population I massive stars in the core-helium burning phase. The initial mass of RSGs is generally considered to be at least $\sim 8 M_{\odot}$, while we \citep[\citetalias{2021ApJ...907...18R} hereafter]{2021ApJ...907...18R} suggested that the lower mass limit can be as small as $7M_{\odot}$ according to the location of RSGs in the color-magnitude diagram, which agrees with the proposal of \citet{2019AandA...629A..91Y}. The radius of RSGs can reach $\sim 1500 R_{\odot}$ \citep{2005ApJ...628..973L}, and they have low surface gravity and high luminosity up to $~3,500-630,000 L_{\odot}$ \citep{2008IAUS..250...97M,2016ApJ...826..224M}.

A complete catalog of RSGs is the basis to study the properties of RSGs, e.g. to constrain the mass range and the evolutionary model of massive stars, and to calibrate the period-luminosity (P-L) relation of RSGs \citep{2006MNRAS.372.1721K,2011ApJ...727...53Y, 2012ApJ...754...35Y, 2018ApJ...859...73S,2019MNRAS.487.4832C, 2019ApJS..241...35R} and the scaling relation between granulation and stellar parameters \citep{2020ApJ...898...24R}. In addition, the statistical study of the complete sample of RSGs is key to estimating the influence of RSGs on a galaxy. For example, the mass loss rates of a complete sample which were determined in M31 and M33 by \citet{2021ApJ...912..112W} illustrate the contribution of interstellar dust by massive stars to the interstellar medium. This is particularly important in that the source of interstellar dust in high-redshift galaxies is still  a puzzle \citep{1975MSRSL...8..369R, 1978A&A....70..227K, 2016ApJ...825...50G}. The luminosity and mass function of a complete sample of RSGs implies the information of the star formation rate (SFR) and the initial mass function (IMF) of a galaxy, in particular at the high-mass end. \citetalias{2021ApJ...907...18R} found that the number of RSGs per stellar mass decreases rapidly with metallicity according to the more-or-less complete sample of RSGs in the Small Magellanic Cloud (SMC), Large Magellanic Cloud (LMC), Triangulum Galaxy (M33) and Andromeda Galaxy (M31), which not only means that the evolution of RSGs depends sensitively on the metallicity (\citealp{1980A&A....90L..17M, 2002ApJS..141...81M, 2013NewAR..57...14M}; \citetalias{2021ApJ...907...18R}), but also implies that the IMF and SFR of galaxies may differ with metallicity.

The complete sample of RSGs in the Milky Way galaxy, our home galaxy, is hardly possible because of our disadvantageous position inside the plane which causes entangling of interstellar extinction and distance and leads to large uncertainty in determining stellar effective temperature and luminosity. Only in the external galaxies is it possible to achieve a complete sample of RSGs under the conditions that the observation is deep enough to cover the faint end and RSGs can be distinguished clearly from other stars. \citet{2007AJ....133.2393M} showed that the contamination by foreground dwarf stars is serious even if a galaxy is not located in the Galactic plane.  Quite some efforts are devoted to distinguish member stars from foreground stars. \citet{2019AandA...629A..91Y,2020AandA...639A.116Y} and \citet{2021AandA...646A.141Y} separated efficiently the SMC and LMC members from foreground stars by using astrometric solution from Gaia/DR2 and combined a variety of color-magnitude diagrams (CMDs) to identify 1,239 and 2,974 RSGs in SMC and LMC, which is a drastic increase from previous studies \citep{1980MNRAS.193..377F, 1981MNRAS.197..385C, 1983ApJ...272...99W, 2000MNRAS.313..271P, 2002ApJS..141...81M, 2003AJ....126.2867M, 2012ApJ...749..177N, 2015AandA...578A...3G, 2011ApJ...727...53Y, 2012ApJ...754...35Y}. This revolutionary progress comes from both more data and the effective method to identify the member stars of the Magellanic Clouds (MCs) since the MC members concentrate on the proper motions expected from the motion of MCs relative to the Galaxy. Unfortunately, this method is not suitable for other galaxies because they are too distant. The main contaminants of the member stars in distant galaxies are faint foreground stars in the CMD (the contaminants and the member stars have similar apparent magnitudes and colors). That means contaminants are mainly distant and faint foreground stars and make it hard to obtain reliable measurements of proper motions and parallaxes from Gaia. For these galaxies, other methods are developed to remove the contamination from the foreground stars. \citet{2016ApJ...826..224M} and \citet{2012ApJ...750...97D} removed foreground stars by radial velocities and spectral type from optical spectroscopy and identified RSGs and membership in M31 and M33 correctly but only for very bright sources. A progress is made by replacing spectroscopy by photometry. \citet{1998ApJ...501..153M}, \citet{2009ApJ...703..420M}, \citet{2012ApJ...750...97D} and \citet{2016ApJ...826..224M} make use of the optical color-color diagram (CCD), $B-V/V-R$, to distinguish foreground dwarfs from RSGs, which is applied to M31 and M33 to find 437 and 749 RSG candidates respectively.
However, as \citet{1998ApJ...501..153M} noted that the difference of RSGs with dwarfs is small in the optical CCD so that very high-accuracy (better than 0.01 mag) photometry is required to distinguish RSGs from dwarfs. Thus, this method is only applicable to bright stars.
This is confirmed by \citetalias{2021ApJ...907...18R}. \citet{2021A&A...647A.167Y} set up a new CCD, i.e. the $r-z/z-H$ diagram to distinguish (super)giants from dwarf stars in NGC 6822 and identified 234 RSG candidates. The difference between the member giant and foreground dwarf stars in the $r-z/z-H$ diagram is greater than in the $B-V/V-R$ diagram. In addition, the longer wavelength of $rzH$ bands in comparison with $BVR$ takes the advantage of RSGs being red so that this method is applicable to fainter stars. The drawback of this method is the requirement of the availability of both near-infrared and optical observations, which certainly lose some objects for distant galaxies.

Recently, we (\citetalias{2021ApJ...907...18R}) establish a new method to remove the foreground dwarf stars. Specifically, the $J-H/H-K$ diagram completely based on the near-infrared photometry, is used to remove the foreground dwarf stars on the basis that the intrinsic color indexes of giant and dwarf stars have clear bifurcations in this CCD \citep{1988PASP..100.1134B}. The underlying physics is that molecules form easier \citep{1995ApJ...445..433A} in dwarfs with higher surface gravity and higher density than giants, causing absorption in the $H$ band and darkening the $H$-band brightness and eventually leading to smaller $J-H$ and bigger $H-K$ than giants. \citet{2021arXiv210807105K} used machine learning algorithms to classify stellar populations including foreground stars in NGC 6822 of which color indexes $J-H$ and $H-K$ are key features for classification. The $J-H/H-K$ diagram only relies on near-infrared photometric data in the $J$, $H$ and $K$ bands which is around the peak emission of RSGs at an effective temperature of 3,000-5,000\,K \citep{2008IAUS..250...97M, 2010ApJ...719.1784N, 2019AandA...629A..91Y}. Moreover, the interstellar extinction is much smaller in near-infrared than in the visual, e.g. $A_{K}$ is about a tenth of $A_{V}$ \citep{2014ApJ...788L..12W}. Mainly using the $J-H/H-K$ criteria which is supplemented by the $r-z/z-H$ diagram and astrometric information from Gaia DR2, \citetalias{2021ApJ...907...18R} gained a complete and pure (i.e. $\sim$ 1\% pollution rate) sample of 5,498 and 3,055 RSGs in M31 and M33 respectively, which increases the number of RSGs in M31 by an order of magnitude and in M33 by a factor of five. The experiment in M31 and M33 proves that the $J-H/H-K$ diagram is an efficient tool to remove the foreground dwarfs and its combination with the $J-K/K$ diagram can identify a complete and pure sample of RSGs as far as the near-infrared observation is deep enough to cover the faint end of RSGs. In comparison, \citet{2021AJ....161...79M} identified 6,412 and 2,858 RSG candidates within the Holmberg radii of M31 and M33 down to $\log L/L_{\odot}=4$ only by their location in the $J-K/K$ diagram which may be polluted by some fraction of foreground dwarf stars.



This work applies the near-infrared CCD (i.e. $J-H/H-K$ diagram) to the Local Group galaxies and attempts to identify the RSGs in them. In addition to the several well-studied galaxies such as LMC, SMC, M31, M33 and NGC 6822, the largest known sample of RSGs in the other Local Group galaxies is currently limited to a dozen \citep{2019A&A...631A..95B}. At present, the number of RSGs is eleven in WLM \citep{2006ApJ...648.1007B,2012AJ....144....2L}, seven in Sextans A
\citep{2014A&A...562A..75B,2015A&A...584A..33B}, two in Sextans B \citep{2019A&A...631A..95B}, eight in IC 1613 \citep{2019MNRAS.483.4751H} and four in IC 10 \citep{2019A&A...631A..95B}, two in NGC 3109 \citep{2007ApJ...659.1198E}, two in the Pegasus Dwarf \citep{2014A&A...562A..75B,2015A&A...584A..33B}, and one in Sagittarius \citep{2018MNRAS.474L..66G}. They are mostly identified by optical spectroscopy, which is time-consuming and applicable only to bright sources. Moreover, spectroscopy may easily confuse red giants with red supergiants. The paper is organized as Section \ref{sec:data} on the data, Section \ref{sec:Extinction correction} on extinction correction, Section \ref{sec:Removing the foreground stars} on the method to remove the foreground stars, and Section \ref{sec:Identifying the RSGs in the CMD} on identification of RSGs followed by Section \ref{sec:summary and conclusion} for summary.

\section{Data and Reduction} \label{sec:data}

The Local Group contains approximately three dozen galaxies, they form a local laboratory for investigating galaxy and stellar evolution, as well as stellar population. A part of these galaxies are investigated in this work by requiring the availability of near-infrared deep observation. On the other hand, they are representative of various types of galaxies to study RSGs in different galactic environments. In details, the sample galaxies include: the dwarf elliptical (dE) galaxy NGC 147 and NGC 185; dwarf spheroidal (dSph) galaxy Pegasus Dwarf (DDO 216), Sextans A and B; dwarf irregular (dIrr) galaxy IC 10, IC 1613, WLM (DDO 221) and Leo A (DDO 69). In addition, SMC (Irr), LMC (SBm) and NGC 6822 (dIrr) that are well studied by \citet{2019AandA...629A..91Y, 2020AandA...639A.116Y, 2021AandA...646A.141Y, 2021A&A...647A.167Y} are also included because the present sample of RSGs can still be improved. The fundamental information about these 12 galaxies is listed in Table \ref{tab:basic information}, including the position, type, diameter, Galactic foreground extinction, and distance modulus ($\mu$) from references.

\subsection{The Dwarf Galaxies: WLM, IC 10, NGC 147, NGC 185, IC 1613, Leo A, Sextans B, Sextans A, NGC 6822, Pegasus Dwarf} \label{sec:data_dwarf}

For all the dwarf galaxies, the $JHK$ photometry is performed for the images taken with the Wide Field Camera (WFCAM) from mid Sep 2005 to Aug 2007 on the 3.8 m United Kingdom Infra-Red Telescope (UKIRT) located in Hawaii \citep{2013ASSP...37..229I}. WFCAM consists of four Rockwell Hawaii-II (HgCdTe $2048 \times 2048$) detectors, each covering $13'.65$ on sky. Most frames work on non-microstep mode with $0.4''/$pixel, some frames work on $2\times2$ microstep mode to give an effective sampling of $0.2''/$pixel\footnote{http://casu.ast.cam.ac.uk/surveys-projects/wfcam/technical/interleaving}. For the images we used in this work, the average seeing on all frames varied between $\sim$ $0.5''-1.5''$.
The images are made available via the WFCAM Science Archive\footnote{http://surveys.roe.ac.uk/wsa} and were reduced by the Cambridge Astronomical Survey Unit (CASU) pipeline\footnote{http://casu.ast.cam.ac.uk/surveys-projects/software-release/fitsio\_cat\_list.f/view}, which produced ASCII-format catalogs with R.A., Decl., magnitude, magnitude error, stellar classification flag, etc. The sources in the $J$, $H$, and $K$ bands are cross-matched and required to align better than $1''$. In each band, the sources with the stellar classification flag of $-1$ (stellar), $-2$ (probably-stellar), and $-7$ (source with bad pixels) are regarded to be point sources. We add an ``N\_Flag" index to the $JHK$ catalogs to indicate the number of bands in which the source is identified as a point source, i.e. 3 means all three bands are labelled "stellar". Sources identified as stellar in at least two photometric bands were taken as initial sample. The data is probably the best from ground-based observation. The process of data reduction and selection is basically the same as that in \citetalias{2021ApJ...907...18R} for M31 and M33, and the related details can be found in \citetalias{2021ApJ...907...18R} and references therein.

\begin{figure}
	\centering
    \includegraphics[scale=0.26]{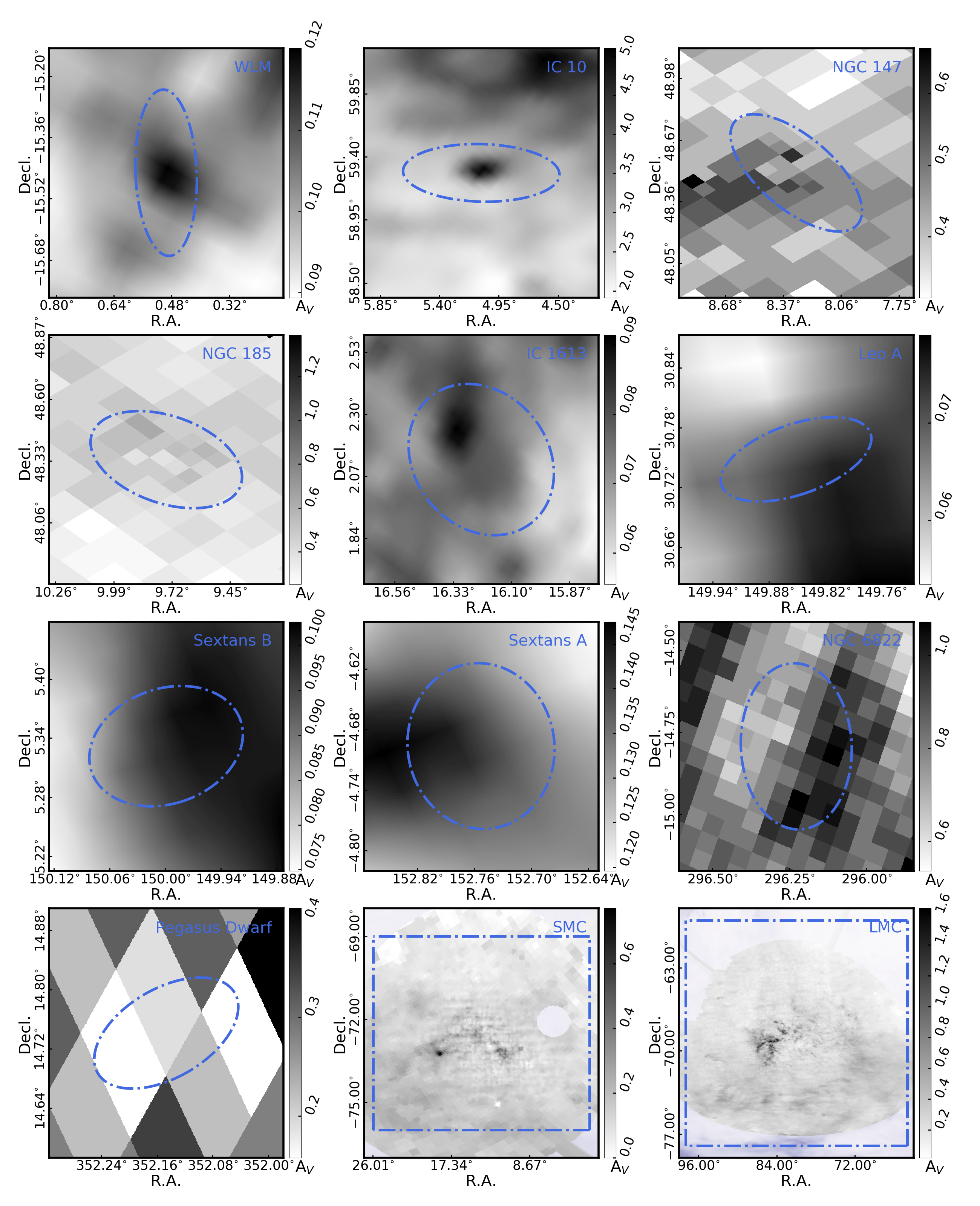}
	\caption{The $V$ band extinction map of all the sample galaxies. The areas enclosed by the royal blue dash-dotted lines are the sky areas of the initial samples. For SMC and LMC, the purple areas indicate that the extinction values are taken from SFD98. \label{fig:dustmaps}}
\end{figure}

For these ten dwarf galaxies, the UKIRT's coverage is far beyond the sky area of the target galaxy. Although the field outside the sky area of the target galaxy may be used as control fields to estimate the pollution rate (\citetalias{2021ApJ...907...18R}), an over large field would bring excessive foreground stars. Therefore, a range of field is set by an ellipse that fits the surface number density of stars based on the initial sample for which the semi-major and semi-minor axes are taken as the radius where the number density drops to the $5\sigma$ level along each axis respectively. Then sources outside this ellipse are removed. The adopted ellipses are shown in Figure \ref{fig:dustmaps}.

\subsection{SMC and LMC} \label{sec:data_MCs}

In the case of SMC and LMC, the $JHK$ magnitudes are taken from the Two Micron All-Sky Survey (2MASS; \citealp{2006AJ....131.1163S}), in which only the sources with photometry quality of ``AAA''\footnote{``AAA'' represents sources with signal-to-noise ratio $>10$ and magnitude uncertainty $<0.10857$ in $JHK$ bands} are selected for the initial sample. The 2MASS instead of the UKIRT/WFCAM results are selected because UKIRT is in the northern hemisphere and the Magellanic Clouds are far south, while 2MASS matches well with the close distance to the Magellanic Clouds.

It should be noted that although the photometry and astrometry of UKIRT are calibrated against 2MASS \citep{2004SPIE.5493..411I,2009MNRAS.394..675H}, its filter system differ from the 2MASS \citep{2009MNRAS.394..675H}, and their photometric magnitudes are different. For the convenience of comparing the results later, the magnitudes of 2MASS are converted into the UKIRT system using the transformation equations in \citet{2009MNRAS.394..675H}.


\section{Extinction correction} \label{sec:Extinction correction}
The Galactic latitudes of the galaxies studied in this work span a large range from $-3^{\circ}$ for IC 10 to $-73^{\circ}$ for WLM, which means a large difference in Galactic foreground interstellar extinction. As shown in Table \ref{tab:basic information}, the foreground extinction $A_{V}$ varies from about 0.1 mag to 4.3 mag, and is smaller than about 0.5 mag for all galaxies except IC 10 (4.3 mag) and NGC 6822 (0.65 mag). Moreover, some galaxies such as LMC and IC 10 have nonnegligible internal extinction. This makes it difficult to compare the results directly from the observed brightness and color index between galaxies. Therefore, the initial sample is corrected for extinction with the two-dimensional (2D) or three-dimensional (3D) dust maps. Unfortunately, there is a lack of the precise measurement of interstellar extinction of both foreground and galaxy itself. Nevertheless, the extinction is not serious for six of the sample galaxies, i.e. IC 1613, Sextans A, Sextans B, Leo A, WLM, Pegasus Dwarf, for which the foreground $A_{K} < 0.02$ and $E(J-K)<0.03$ approximately and no serious internal extinction is reported. Thus the results about these galaxies should be little influenced by the extinction correction. Under the available extinction maps, the extinction is corrected accordingly for different galaxies as follows.

\begin{itemize}
  \item NGC 6822, NGC 147, NGC 185, and Pegasus Dwarf. The foreground reddening value $E(B-V)$ of these four galaxies are retrieved from the 3D extinction maps by \citet{2019ApJ...887...93G} (hereafter Bayestar19 extinction map) based on Gaia parallaxes and stellar photometry from Pan-STARRS 1 and 2MASS. The distance of all sources is set to 50 kpc in retrieving the reddening values to include all the foreground extinction. Although this may overestimate the extinction of the nearby foreground stars, which means this fraction of stars will be corrected to be bluer than the true value, they would still be considered as foreground stars in the CCD which will be shown later. This reddening value is reasonable for the member stars of the galaxy as they must have experienced the full foreground extinction of the Galaxy.
  \item IC 10, IC 1613, Sextans A, Sextans B, WLM, and Leo A. For these galaxies, the SFD98 2D extinction map \citep{1998ApJ...500..525S} is used to correct the extinction.  Among them, the Bayestar19 extinction map is unavailable for IC 1613, Sextans A, Sextans B, WLM, and Leo A.  About IC 10, it is a starburst galaxy, thus dust-rich. The SFD98 other than the Bayestar19 extinction map is chosen. The reason is that the SFD98 extinction map is constructed from the dust far-infrared emission and further converted to $E(B-V)$, so the reddening value in SFD98 actually includes both the host galaxy's internal extinction and the Galactic foreground extinction, a feature that is beneficial for correcting the extinction of IC 10, though the SFD98 map may suffer some uncertainty in the amount of extinction and inadequate resolution. In comparison, the Bayestar19 extinction only takes the foreground extinction into account.
  \item LMC and SMC. The foreground extinction $A_{V}$ is only 0.2 and 0.1 mag for LMC and SMC respectively and negligible, but they have significant star-formation regions and the internal extinction has to be considered. Most of the reddening value $E(V-I)$ is taken from the 2D extinction map based on red clump stars by \citet{2021ApJS..252...23S} (hereafter OGLE extinction map).  Then $E(V-I)$ values from OGLE reddening maps are converted into $E(B-V)$ using $E(B-V)=E(V-I)/1.237$ \citep{2011ApJ...737..103S}. The OGLE reddening map is preferred over the SFD98 map in the overlap region because the dust temperature structure is not sufficiently resolved leading to the unreliable reddening values from SFD98 for MCs \citep{1998ApJ...500..525S}.
\end{itemize}

The adopted extinction maps of the 12 galaxies are displayed in Figure \ref{fig:dustmaps}, and the $E(B-V)$ from reddening maps are converted into the extinction value in the UKIRT $JHK$ and the Panoramic Survey Telescope And Rapid Response System (PS1) $grizy$ bands by using the extinction law from \citet{2019ApJ...877..116W}, and the apparent magnitude after extinction correction is calculated for the initial samples.

\section{Removing the foreground stars} \label{sec:Removing the foreground stars}

\subsection{The Gaia Criteria} \label{sec:Gaia Criteria}

Before using the $(J-H)_{0}\ vs.\ (H-K)_{0}$ and $(r-z)_{0}\ vs.\ (z-H)_{0}$ CCDs, the astrometric information from the Gaia early data release 3 (EDR3) is taken to remove some foreground dwarf stars to suppress the dispersion in the CCDs. Because the astrometric measurements depend on no astrophysical model, this step would not induce additional error. The proper motions of the SMC and LMC objects are mostly measured in the sample so that they can be applied directly to select the candidates for the galaxy's members. Meanwhile, such measurements are unavailable for the stars in other much more distant (by a 5-6 magnitude larger $\mu$) galaxies, therefore, the Gaia measurements are used to remove only a small fraction of the foreground stars other than to choose the members. Thus, this criteria are designed into two types for the dwarf galaxies and the MCs respectively.

\subsubsection{The Dwarf Galaxies} \label{sec:Gaia Dwarf Galaxies}

For the dwarf galaxies, a star is considered to be a foreground object if it satisfies either the parallax or the proper motion constraints with reliable astrometric solution, i.e. the relative error is smaller than 20\% (i.e. $|\sigma_{\omega}/\omega|$, $|\sigma_{\mu_{\alpha*}}/\mu_{\alpha*}|$ and $|\sigma_{\mu_{\delta}}/\mu_{\delta}|$ are all smaller than 20\%).
Specifically, the sources with reliable distances less than the Milky Way scale  (i.e. 50 kpc; \citealt{2017RAA....17...96L}) are removed. The distances are calculated with the Smith-Eichhorn correction method from the Gaia-measured parallax and its error \citep{1996MNRAS.281..211S}. Besides, we remove the sources with reliable proper motion greater than that expected for a star with a velocity of 500\ km/s at the distance of $\sim800\ \mathrm{kpc}$ typical for these dwarf galaxies \citep{2019ApJ...872...24V}, corresponding to 0.2 mas/yr:
\begin{equation} \label{eq:pm_ra}
	\left| \mu_{\alpha*} \right| > 0.2 \mathrm{mas\ yr^{-1}} + 2.0\sigma_{\mu_{\alpha*}},
\end{equation}
\begin{equation} \label{eq:pm_dec}
	\left| \mu_{\delta} \right| > 0.2 \mathrm{mas\ yr^{-1}} + 2.0\sigma_{\mu_{\delta}},
\end{equation}
where $\mu_{\alpha*}$ and $\mu_{\delta}$ are the proper motions in right ascension and declination respectively. The criteria are the same as that applied in \citetalias{2021ApJ...907...18R} for M31 and M33.

\subsubsection{The Magellanic Clouds} \label{sec:Gaia Magellanic Clouds}

\begin{figure}
	\centering
    \includegraphics[scale=0.5]{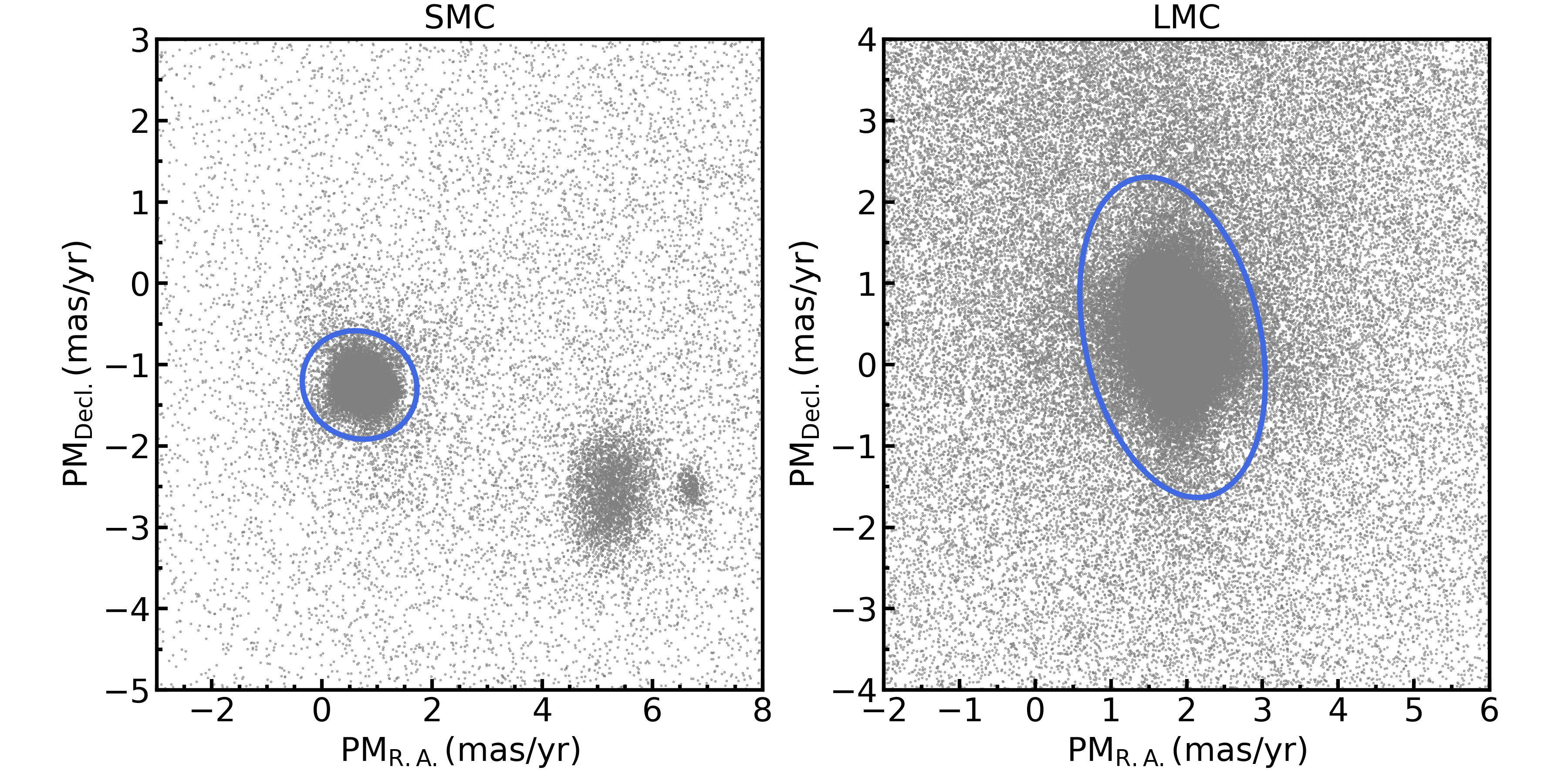}
	\caption{Two-dimensional distribution of $\mathrm{PM_{R.A.}}$ vs. $\mathrm{PM_{Decl.}}$ for the initial samples of  SMC (left) and LMC (right). The royal blue ellipse marks the $5\sigma$ level of the proper motions. \label{fig:pm_mcs}}
\end{figure}

\citet{2019AandA...629A..91Y} and \citet{2020AandA...639A.116Y} already find that the proper motions of the stars in SMC and LMC well concentrate around the overall motion of the galaxy even with the Gaia/DR2 data, which is used to identify the members. In the Gaia/EDR3, the proper motions of the stars in our selected sky areas of SMC and LMC are measured for about 99\% sources in the initial sample, specifically 132,581/133,843 for SMC and 768,192/776,831 for LMC. With many more and higher-accuracy measurements in Gaia/EDR3, the concentration becomes more compact. We make use of this information to exclude the foreground stars instead of identifying the members, i.e. the star is considered as a foreground source if its proper motions differ significantly at $>5\sigma$ level from the systematic proper motions of the SMC and LMC. In detail, an ellipse is taken to fit the number density distribution profile of $\mathrm{PM_{R.A.}}$ ($\mu_{\alpha*}$) vs. $\mathrm{PM_{Decl.}}$ ($\mu_{\delta}$) from the initial sample as shown in Figure \ref{fig:pm_mcs}, and the semi-major axis and semi-minor axis are five times the dispersion of surface density along each axis respectively, i.e. $5\sigma$. The resultant geometric parameters are listed in Table \ref{tab:ellipses}.
The sources that locate outside the $5\sigma$ level ellipse and whose proper motion errors are less than 20\% ($|\sigma_{\mu_{\alpha*}}/\mu_{\alpha*}|$ and $|\sigma_{\mu_{\delta}}/\mu_{\delta}|$ are both smaller than 20\%) are regarded as the foreground stars and removed. Considering the closer distance of the MCs and also the physical scale of the MCs, the distance separation between the foreground and member stars is set closer, i.e. 35 kpc. That is, the sources with $|\sigma_{\omega}/\omega|$ less than 20\% and whose distances are less than 35 kpc are also removed.

\subsection{The $(J-H)_{0}\ vs.\ (H-K)_{0}$ Diagram Criteria} \label{sec:JHK Diagram}

\subsubsection{Dwarf Galaxies} \label{sec:JHK Dwarf Galaxies}

\begin{figure}
	\centering
    \includegraphics[scale=0.24]{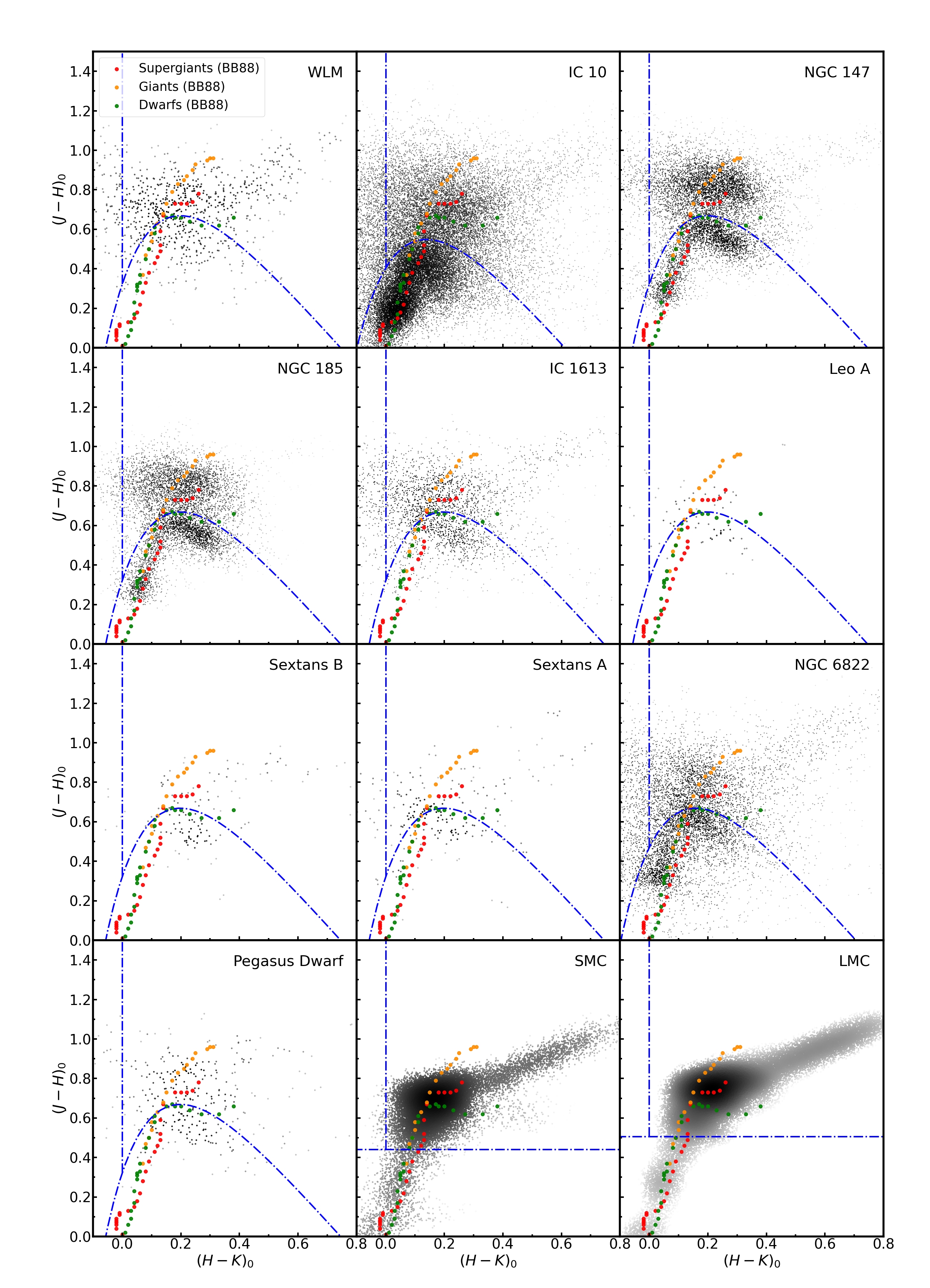}
	\caption{The $(J-H)_{0}\ vs.\ (H-K)_{0}$ diagram of the good-quality measurements with ``N\_Flag=3" and the $JHK$ photometric errors less than 0.20 mag for ten dwarf galaxies, and the 2MASS ``AAA'' sources for SMC and LMC. In addition, the sources that are more than one magnitude fainter than TRGB or considered to be foreground stars by the Gaia criteria are discarded in this figure. The colors represent the surface density that decrease from dark gray to light gray. The criteria to remove the foreground dwarfs are shown by dash-dotted lines, which is compared with the intrinsic color indexes of dwarfs (green dots), giants (yellow dots) and supergiants (red dots) by \citet[BB88 for short]{1988PASP..100.1134B}. \label{fig:UKIRT_Criteria}}
\end{figure}

After removing foreground stars with the Gaia criteria, the $(J-H)_{0}\ vs.\ (H-K)_{0}$ diagrams for the reserved sources with N\_Flag=3 and $\sigma_{JHK}<0.2$ are displayed in Fig \ref{fig:UKIRT_Criteria} for all the 12 galaxies, where the sources 1 mag fainter than Tip-RGB (TRGB) (see Section \ref{sec:TRGB}) are excluded because their photometry error is relatively large to cause extra confusion.

Instead of defining the borderline galaxy by galaxy, a uniform function is used to separate giants and supergiants from dwarfs for the ten dwarf galaxies. The locations in the $(J-H)_{0}\ vs.\ (H-K)_{0}$  diagram of the foreground stars are expected to be the same because these stars are Galactic and metallicity affects little on the near-infrared colors (see e.g. \citealt{2017AJ....153....5J}). Thus the borderline to enclose the dwarfs should be uniform independent of the target galaxy. Moreover, some galaxies are so small that there are not enough foreground stars to delineate the dwarf branch. In practice, the dwarf stars in the sightlines towards NGC 147 and NGC 185 are chosen to define the borderline with the functional form as following:
\begin{equation} \label{eq:UKIRT_Criteria}
	(J-H)_{0} = -\exp(-6.50[(H-K)_{0} + 0.01]) - 1.68(H-K)_{0} + 1.26.
\end{equation}

However, this borderline does not suit for IC 10 and NGC 6822. This may be caused by inaccurate extinction correction. As shown in Table \ref{tab:basic information}, they have the largest foreground extinction, and the present extinction map may have not enough spatial resolution to account for the extinction variation, which is particularly so for IC 10. One more possible reason is that these two galaxies are metal-poor in comparison with the Galactic disk (see Section \ref{sec:TRGB}) in which the stars are bluer and shifted to the dwarf stars region. No matter what the reason may be, the borderline is adjusted separately according to the measured $(J-H)_{0}\ vs.\ (H-K)_{0}$ by eye-check, which shifts the line defined by Eq. \ref{eq:UKIRT_Criteria} leftward by 0.06 mag and downward by 0.12 mag for IC 10, and 0.04 mag leftward for NGC 6822.

\subsubsection{The Magellanic Clouds} \label{sec:JHK Magellanic Clouds}

Although LMC and SMC are as metal-poor as IC 10 and NGC 6822, the problem of the mixing of the bluer member giant stars with the foreground dwarf stars is highly alleviated by the removal of foreground stars from their proper motions and parallaxes done in Section \ref{sec:Gaia Magellanic Clouds}. In order to take the metallicity effect into account, an independent borderline is determined for them. As displayed in Figure \ref{fig:UKIRT_Criteria}, we only take a constant cutoff in $(J-H)_{0}$ with the specific value of 0.441 and 0.505 for SMC and LMC respectively. It should be kept in mind that this cutoff is surely bluer than some foreground dwarf stars, which means the member stars may be polluted by foreground dwarf stars. Though this is not perfect, the near-infrared criteria removes $\sim 14\% $ and $\sim 20\%$ foreground stars additionally after the Gaia criteria (see Table \ref{tab:numbers} for details).

With the borderline determined above, $(H-K)_{0}<0$ is required for all the galaxies to remove the stars certainly bluer than RSGs.

\subsection{The $(r-z)_{0}\ vs.\ (z-H)_{0}$ Diagram Criteria} \label{sec:rzH Diagram}

Actually the $(J-H)_{0}\ vs.\ (H-K)_{0}$ diagram works very efficiently and is highly consistent with the $(r-z)_{0}\ vs.\ (z-H)_{0}$ criteria (\citetalias{2021ApJ...907...18R}). We introduce the $(r-z)_{0}\ vs.\ (z-H)_{0}$ diagram as an auxiliary tool that can be helpful to further remove some foreground stars which are very close to the borderline in the $(J-H)_{0}\ vs.\ (H-K)_{0}$ diagrams but may be significantly distinguishable in the CCD $(r-z)_{0}\ vs.\ (z-H)_{0}$ especially for some relatively blue stars.

\begin{figure}
	\centering
    \includegraphics[scale=0.26]{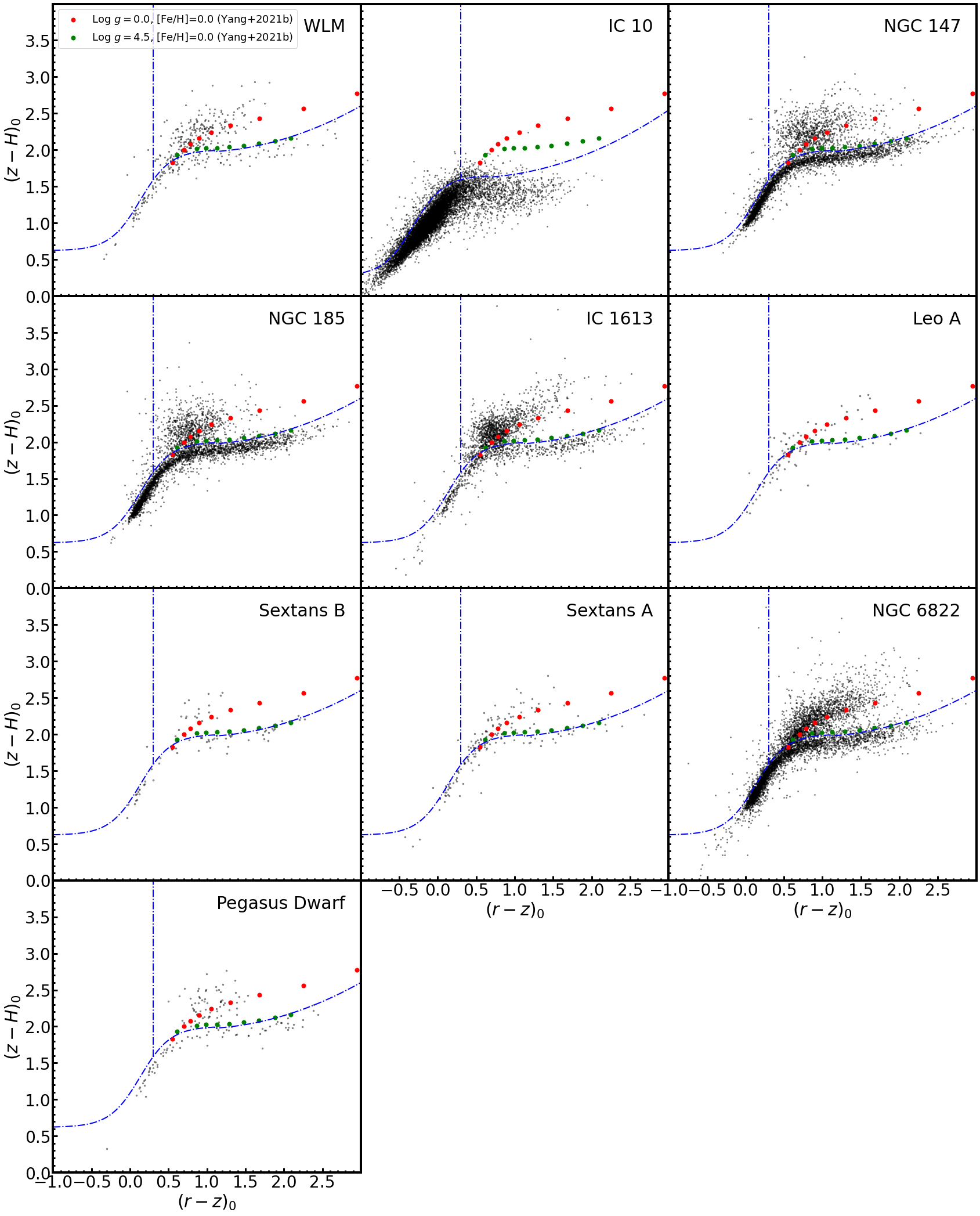}
	\caption{The $(r-z)_{0}\ vs.\ (z-H)_{0}$ diagram of the good-quality measurements with “N\_Flag=3” and the $rzH$ photometric errors less than 0.1 mag. The symbol convention follows Figure \ref{fig:UKIRT_Criteria}. The intrinsic color indexes of dwarfs and supergiants from \citet[Yang+2021b for short]{2021A&A...647A.167Y} are shown for comparison. \label{fig:PS1_Criteria}}
\end{figure}

The optical photometry in the $r$ and $z$ band is taken from the PS1/DR2 catalog for all but the Magellanic Cloud galaxies limited by its coverage. The forced mean PSF magnitude in the catalog is adopted for its consideration of both photometric depth and accuracy. The data flags of PS1/DR2 are complex and only those with good measurements are taken into account. The UKIRT and PS1 catalogs are cross-matched by a radius of $1''$. The functional form and coefficients of the dividing lines between dwarfs and giants on the $(r-z)_{0}\ vs.\ (z-H)_{0}$ diagrams for the 10 dwarf galaxies are the same as those given in \citetalias{2021ApJ...907...18R} for M33. For IC 10, the dividing line is adjusted separately, with the original dividing line shifted leftward by 0.45 mag and downward by 0.35 mag. As shown in Figure \ref{fig:PS1_Criteria}, sources falling below the dividing line or $(r-z)_{0}<0.3$ are considered to be foreground dwarf stars. The criteria are applied to the PS1 sources with ``good measurement" and the photometric error in the $r$, $z$, $H$ bands is less than 0.1 mag. This criteria are not applied to LMC and SMC due to PS1 data unavailable for them.


The total and individual numbers of foreground stars removed by the above three methods are summarized in Table \ref{tab:numbers} as well as the number of the reserved member stars in each galaxy. It can be seen that the near-infrared CCD is highly efficient to remove the foreground dwarf stars. For the ten dwarf galaxies, the NIR CCD removes mostly more than 95\% of dwarf stars. In comparison, the Gaia criteria removes generally less than 10\%, which can be explained by the far distances around 800 kpc of these galaxies so that many of the objects are not measurable by Gaia.  For the MCs, the situation changes. As mentioned above, 99\% of the stars in the sky area of the MCs have the astrometric information of Gaia, and the priority is given to the Gaia proper motion and parallax to remove the foreground stars. Consequently, the Gaia criteria removes 67\% and 84\% of the foreground stars for SMC and LMC respectively, while still the NIR CCD removes 58\% and 76\%, comparably efficient to the Gaia criteria. More important is that the NIR CCD removes 9,738 and 59,875 sources that are not removed by the Gaia criteria, which means the use of the Gaia criteria alone cannot make up a pure sample.

\section{Identification of RSGs in the $(J-K)_{0}\ vs.\ K_{0}$ Diagram} \label{sec:Identifying the RSGs in the CMD}

\subsection{Tip-RGB} \label{sec:TRGB}

The study of M31 and M33 by \citetalias{2021ApJ...907...18R} proves that the position in the color-magnitude diagram $(J-K)_{0}\ vs.\ K_{0}$ of TRGB is a key point to mark the faint end of $K_{0}$ of RSGs and the dividing line of color index $(J-K)_{0}$ between RSG and AGB stars. Referring to the Padova stellar evolutionary track, the criterion for a star to be a RSG is that this star must not evolve through the AGB phase. This criterion changes the lower limit of the initial stellar mass of RSG with metallicity. At [M/H]=$-1.5$, this limit is about 5 $M_{\odot}$, and goes to about 8 $M_{\odot}$ at [M/H] = 1.0. Besides, the lower limit of the $K$-band brightness also increases with metallicity. At the same time, the position of TRGB changes in the same way, i.e. the brightness of TRGB becomes bright with metallicity. Consequently, the lower limit of RSG always coincides with the $K$-band brightness of TRGB despite of the variation of metallicity. So this work complies with \citetalias{2021ApJ...907...18R}, also taking the $K_{0}$ magnitude of TRGB as the fainter limit of RSGs, which means that RSGs should be brighter than TRGB in the $K$ band.

After removing the foreground stars, the photometry is deep enough to cover the TRGB in the $(J-K)_{0}\ vs.\ K_{0}$ diagram for all but Sextans A and Sextans B galaxies in the sample. \citetalias{2021ApJ...907...18R} used the saddle point in the $(J-K)_{0}\ vs.\ K_{0}$ diagram to determine the position of TRGB, i.e. the minimum density along the $K_{0}$ axis while the maximum density along the $(J-K)_{0}$ axis. This saddle point originates from the combining effect of a few factors, i.e. the very quick evolution in TRGB as the dominant factor, and the increasing number of RGB stars and the decreasing completeness with decreasing magnitude. As a general result, the density of stars  in the $(J-K)_{0}\ vs.\ K_{0}$ diagram increases from the bottom gradually, then decreases up to TRGB  and increases again to AGB, where TRGB is at the saddle point. However, the feature of saddle point is not obvious for some galaxies in the sample because there are not enough AGBs, which occurs usually in small galaxies. In order to enable a more general determination of the TRGB position, the Poisson Noise weighted star counts difference in two adjacent bins \citep[hereafter the PN method]{2018AJ....156..278G} is used to detect TRGB. \citet{2002AJ....124.3222B} points out that TRGB has a certain width in brightness, which leads to that the PN-detected TRGB would be fainter than the true TRGB. We thus consider the PN-detected $K_{0}$ magnitude to be the faint end of TRGB, while the bright end of TRGB is taken to be the first brighter local minimum of the PN filter, and the average of the bright and faint ends as the true magnitude of TRGB. The result is checked visually to exhibit no systematic deviation.

\begin{figure}
	\centering
    \includegraphics[scale=0.5]{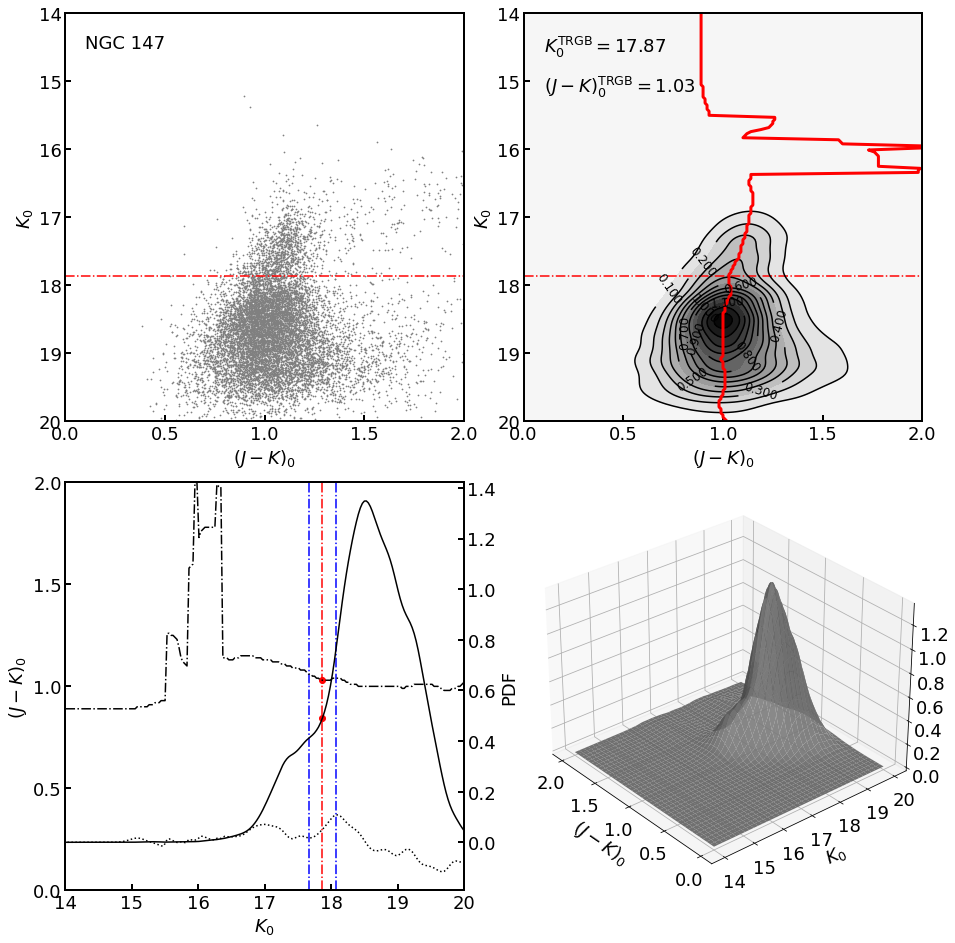}
	\caption{NGC 147 is taken as an example to illustrate how TRGB is determined. Upper left: the CMD of the member stars with the red dash-dotted line indicating the position of the TRGB; Upper right: the probability density function (PDF) of the member stars, where the red solid line shows the $K_{0}\ vs.\ (J-K)_{0}$ relation of the ridge line; Lower left: The black dash-dotted line shows the $\mathrm{PDF}_\mathrm{ridge}\ vs.\ (J-K)_{0}$ relation, the black solid line shows the $\mathrm{PDF}_\mathrm{ridge}\ vs.\ K_{0}$ relation, and the black dotted line shows the response of the PN filter. The blue dash-dotted lines indicate the bright and faint ends of the TRGB, respectively, and the red dash-dotted lines indicate the true position of TRGB; Lower right: the 3D density distribution of the member stars. \label{fig:TRGB_N147}}
\end{figure}

Algorithmically, the number density distribution of the member stars in the $(J-K)_{0}\ vs.\ K_{0}$ diagram is calculated using kernel density estimation by a Gaussian kernel with a width of 0.1 and can be expressed as $\mathrm{PDF}_\mathrm{stars} = f(K_{0},(J-K)_{0})$. The 3D density distribution $\mathrm{PDF}_\mathrm{stars}$ is subsequently sliced along the $K_{0}$-axis in steps of 0.01, and the highest point is taken as the maximum value in each slice. In this way, the ridge of the 3D density distribution can be determined, and the ridge is then projected along the directions parallel to $(J-K)_{0}$ and parallel to $K_{0}$ to obtain the relations $\mathrm{PDF}_\mathrm{ridge} (K_{0})$ and $\mathrm{PDF}_\mathrm{ridge} ((J-K)_{0})$, respectively. Afterwards the PN filter is used to detect the edge of the $\mathrm{PDF}_\mathrm{ridge} (K_{0})$ curve, and the averaged bright and faint ends is taken as the TRGB magnitude. Meanwhile, the color index $(J-K)_{0}$ of TRGB is determined by the $\mathrm{PDF}_\mathrm{ridge}\ (K_{0})\ vs.\ (J-K)_{0}$ relation. The procedure for NGC 147 is shown in Figure \ref{fig:TRGB_N147} as the example.

For Sextans A and Sextans B, the TRGB is not identifiable by the PN method or by eyes. Considering $\mu$ determined previously being around 25.8 mag \citep{2012AJ....144....4M}, very probably the photometry is not deep enough to cover the TRGB. Instead, the observed $(J-K)_{0}/K_{0}$ CMD is matched with the model to infer the $\mu$ and then the position of the TRGB. By taking  $\mu$ as an independent variable in the fitting, the observed and model CMDs are divided into subregions with a grid size of $0.01 \times 0.01$, and $\mu$ is obtained by minimizing the $\chi^{2}$, where $\chi^{2}$ is expressed as following,
\begin{equation} \label{eq:CMD_model}
	\chi^{2}=\sum_{i} \frac{\left(N_{D_{i}}-N_{M_{i}}\right)^{2}}{N_{D_{i}}},
\end{equation}
where $N_{D_{i}}$ and $N_{M_{i}}$ is the number of stars in the $i$th subregion of the observed and model CMD respectively, and the details of CMD fitting can be found in \citet{2001ApJS..136...25H}. The CMD model is constructed from a series of isochrones\footnote{\href{http://stev.oapd.inaf.it/cgi-bin/cmd}{http://stev.oapd.inaf.it/cgi-bin/cmd}} with $\log\mathrm{(age/yr)}$ from 6 to 9.9 in a step of 0.05 based on the Padova stellar evolutionary tracks \citep{2012MNRAS.427..127B,2013MNRAS.434..488M,2014MNRAS.445.4287T,2014MNRAS.444.2525C,2015MNRAS.452.1068C,2017ApJ...835...77M,2019MNRAS.485.5666P,2020MNRAS.498.3283P}. Then each isochrone point is assigned an absolute number of stars per unit mass based on initial mass function \citep{2001MNRAS.322..231K,2002Sci...295...82K}. The circumstellar dust of oxygen-rich AGBs stars is neglected in the model, and the composition of circumstellar dust of carbon-rich AGBs is chosen as graphite \citep{1998A&A...332..135B}. The metallicity [M/H] of the model is set to $-$1.4\footnote{Here we simply take [M/H]$\sim$[Fe/H].} according to the measurements by \citet{2003AJ....126..187D} and \citet{2005AJ....130.1558K}.

\begin{figure}
	\centering
    \includegraphics[scale=0.5]{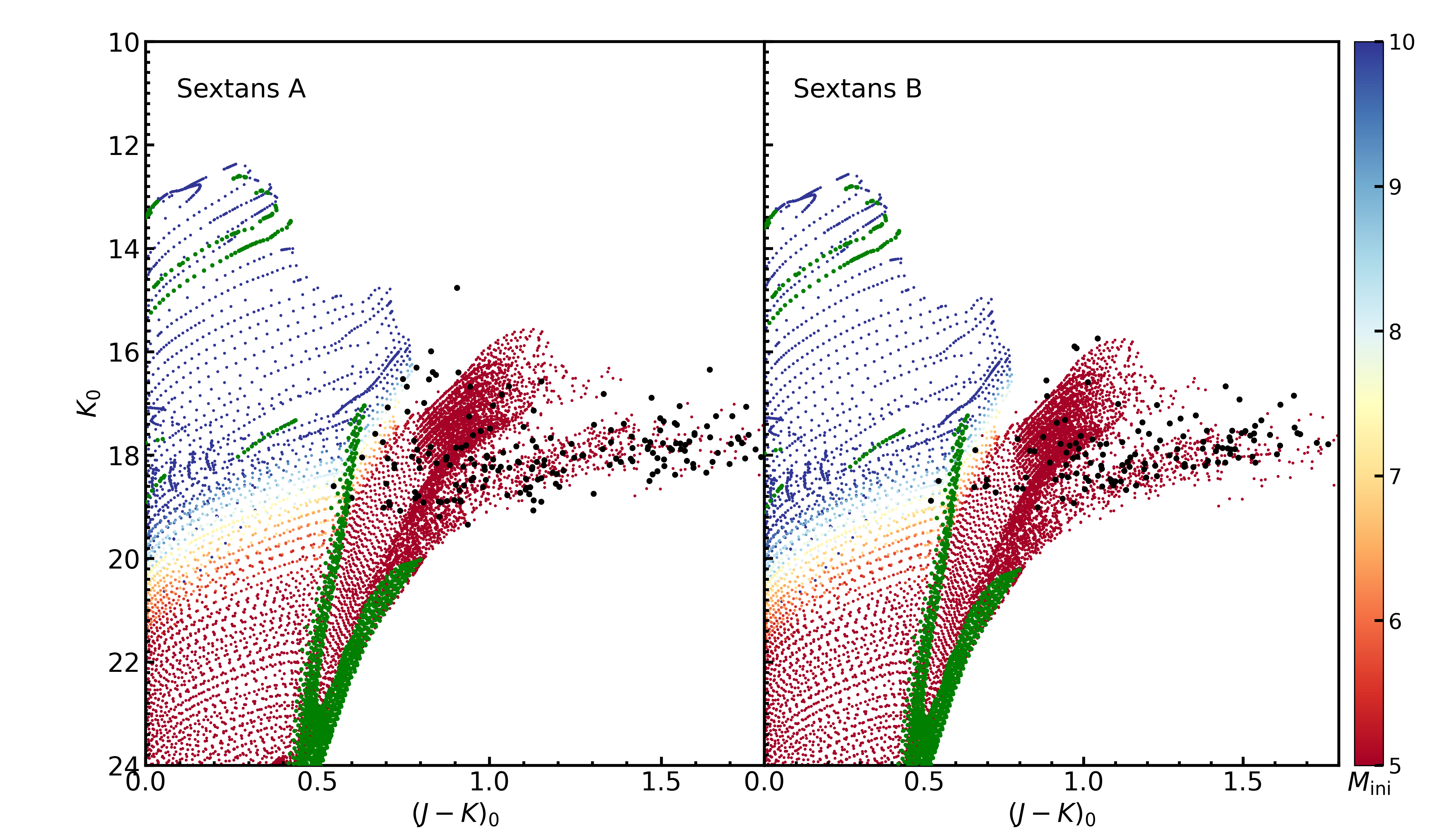}
	\caption{The best Padova model fitting for the observed CMD of Sextans A (left) and Sextans B (right). The black dots are member stars in this work. The colored dots are isochrones with the color representing the initial stellar mass, and the green dots represent RGB stars (or quick stage of red giant for intermediate and massive stars) marked in the Padova isochrones. The reddest RGB star in the Padova isochrones is regarded as TRGB.
	\label{fig:CMD_fitting}}
\end{figure}

The best results of the Sextans A and Sextans B CMD fitting are shown in Figure \ref{fig:CMD_fitting}, where the $\mu$ turns out to be 25.95 and 26.15 respectively. In comparison, this value is about 0.1-0.2 mag bigger than previous measurements of a distance between 1.3 to 1.5 Mpc, i.e. $\mu$ from 25.6-25.9 mag, but still comparable. For other galaxies, $\mu$ is calculated directly from the difference of observed $K^{\rm TRGB}$  with the absolute magnitude of TRGB determined from $(J-K)_{0}$ after taking the metallicity effect into account according to \citet{2018AJ....156..278G} and \citetalias{2021ApJ...907...18R}:
\begin{equation} \label{eq:TRGB_mag}
	M_{K}^\mathrm{TRGB} = -2.78[(J-K)_{0}-1.0]-6.26.
\end{equation}

The magnitude $K_{0}$ and color index $(J-K)_{0}$ of TRGB, and the $\mu$ for the 12 galaxies are listed in Table \ref{tab:parameters of galaxies}. Most of the $\mu$ derived in this work are consistent with the results in the literatures with a dispersion of around 0.2-0.3 mag. However, it deserves to mention that the derived $\mu$ of SMC and LMC, 18.26 and 18.72 respectively is about 0.2 mag smaller than the highly consistent values of 18.50 and 18.90 respectively. Since the two galaxies are the closest and the observational data is of excellent quality, the observed position of TRGB in the CMD should be quite reliable. Indeed, the measured $I_0$ of TRGB in LMC by \citet{2000AJ....119.1197S}, 14.54, is in excellent agreement with our $K_0=12.00$ with the color index $(I-K)_0 \sim 2.5$ for TRGB \citep{2004A&A...424..199B}. The difference in $\mu$ between SMC and LMC, 0.46, also coincides very well with the result of \citet{2020ApJ...891...57F}. Thus the discrepancy must be caused by the conversion from $(J-K)_0$ to $M_{K}$ \citep{2018AJ....156..278G}, which may imply that this relation needs further examination in future work. As a result, it should be kept in mind that $\mu$ may be systematically smaller than the true value in such case.

\subsection{The Sample of RSGs} \label{sec:Sample of RSGs}

\subsubsection{RSGs} \label{sec:RSGs}
\begin{figure}
	\centering
    \includegraphics[scale=0.26]{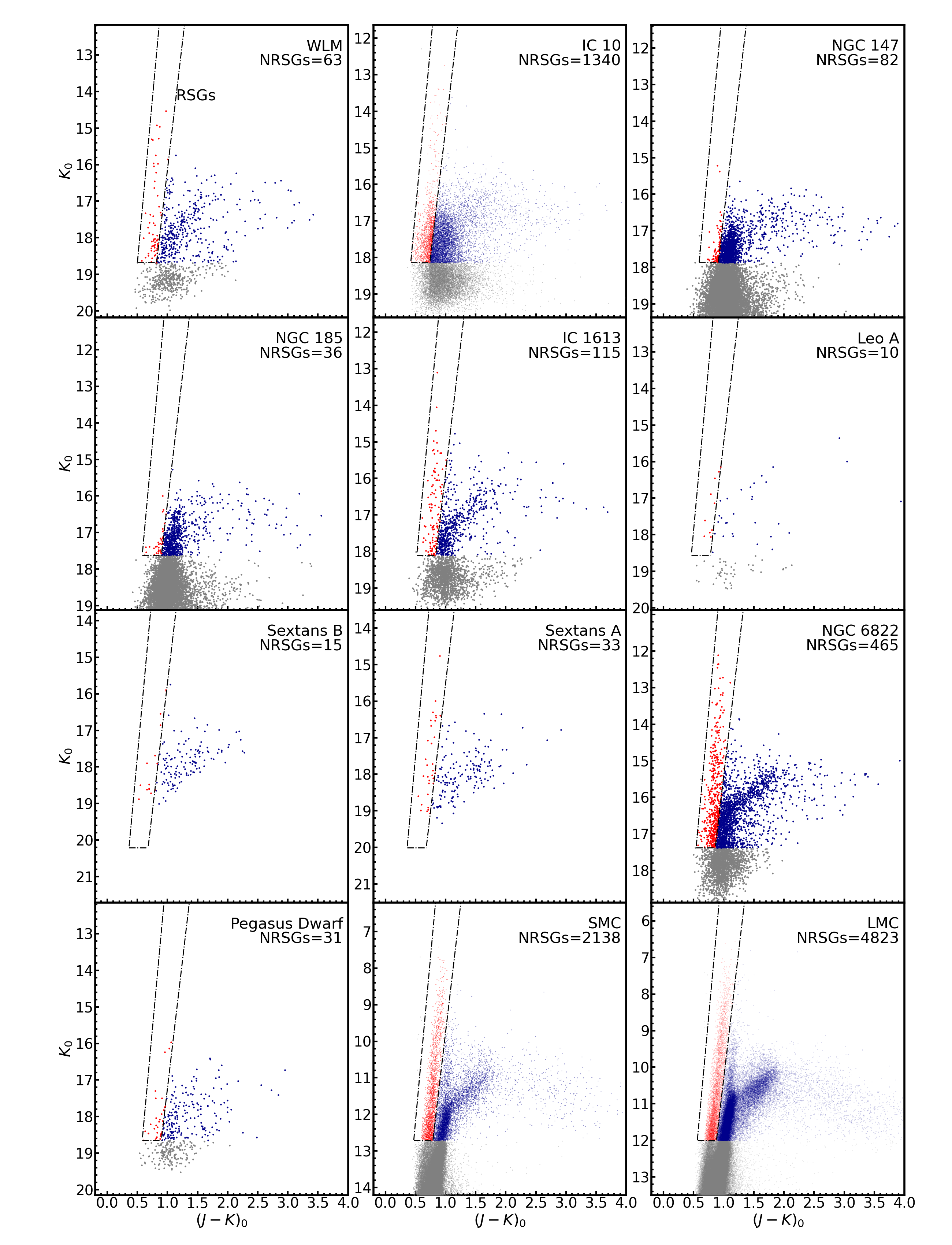}
	\caption{The color-magnitude diagram of the member stars for each galaxy. The RSGs and AGBs are shown in red and dark-blue while other member stars are shown in gray, while the dash-dotted lines define the region of RSGs. Also displayed is the number of RSGs in the galaxy. \label{fig:fgd_member}}
\end{figure}

The RSG branch is very obvious in the $J-K/K$ diagram of member stars for the galaxies with numerous objects as shown in Figure \ref{fig:fgd_member} so that their region can be directly defined. On the other hand, this branch becomes fuzzy for the galaxies with not so many RSG stars. Instead of defining the RSG region for each galaxy individually, we take the most accurately measured one, i.e. the LMC as the standard to mark the red and blue boundaries of RSGs. This boundary is then shifted by the differences in the TRGB color $(J-K)_{0}$ and the brightness $K_{0}$ for other galaxies, i.e. $\Delta(J-K)_0=(J-K)^\mathrm{TRGB}_{0,\mathrm{galaxy}}-(J-K)^\mathrm{TRGB}_{0,\mathrm{LMC}}$ and $\Delta K_0=K^\mathrm{TRGB}_{0,\mathrm{galaxy}}-K^\mathrm{TRGB}_{0,\mathrm{LMC}}$. Analytically, the boundaries take the following form:
\begin{equation} \label{eq:blue boundaries}
	\mathrm{blue\ boundary}: K_{0}=-18.00(J-K)_{0}+24.10+18.00\Delta(J-K)_0+\Delta K_0, \\
\end{equation}
\begin{equation} \label{eq:red boundaries}
	\mathrm{red\ boundary}: K_{0}=-14.00(J-K)_{0}+24.30+14.00\Delta(J-K)_0+\Delta K_0.
\end{equation}

Because all the ten dwarf galaxies are much more distant by about 5 mag larger $\mu$ than LMC and SMC, the photometric error in the NIR bands increases apparently, thus the blue boundaries of these 10 dwarf galaxies are shifted to the left by 0.11 mag to account for the broadening of the RSG branch caused by the photometric error. One potential uncertainty of this method is that the shape of RSG branch may differ with metallicity. A rough examination by the Padova model finds very small change: when the metallicity increases 2.5 dex ([M/H] from $-2.0$ to 0.5), the width ($J-K$) of the RSG branch increases by about 0.03 mag. The visual check of the $J-K/K$ diagram looks fine, and no further correction is performed.

\subsubsection{Completeness and pureness} \label{sec:Completeness and pureness}

\begin{figure}
	\centering
    \includegraphics[scale=0.5]{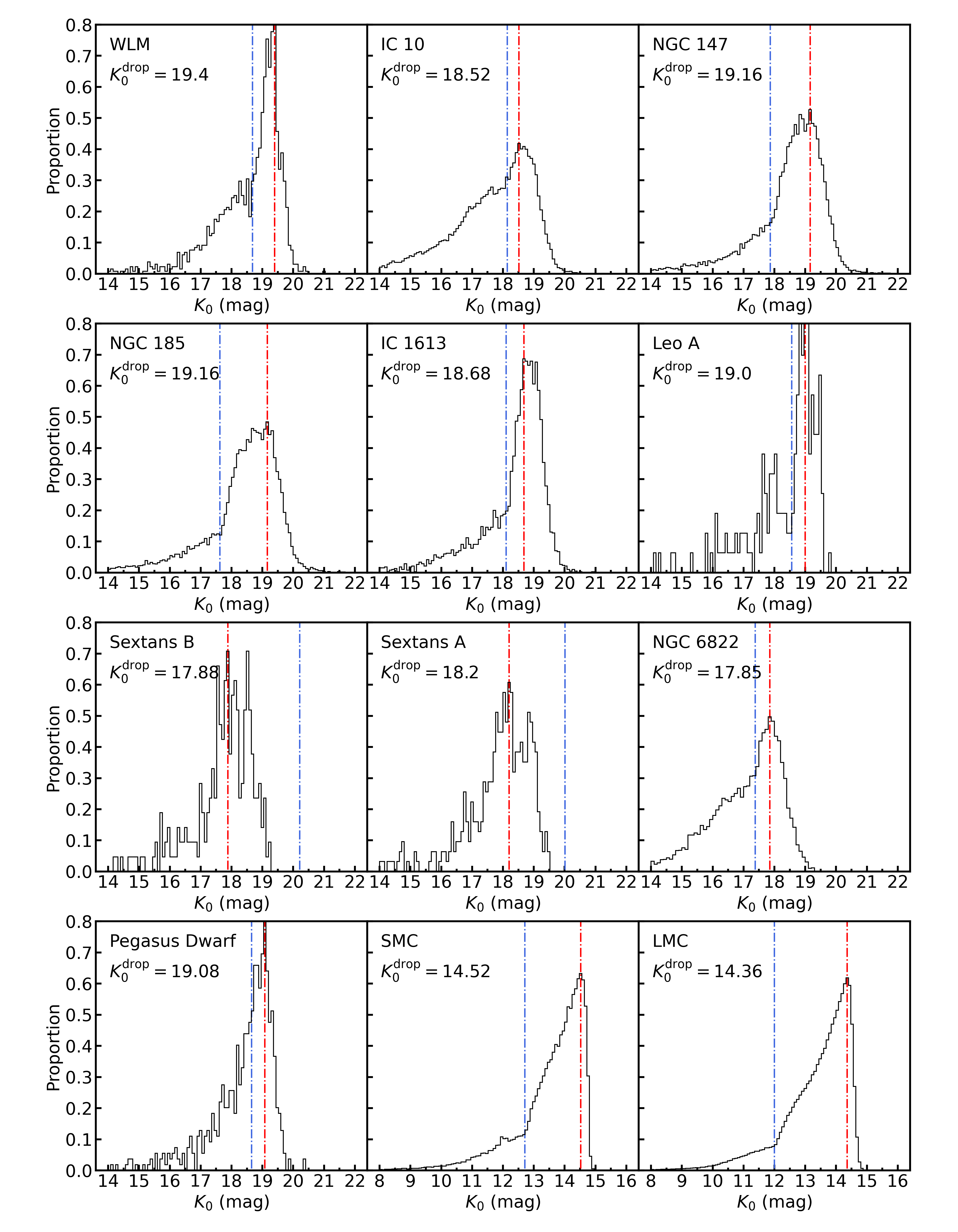}
	\caption{The distribution of the observational $K_{0}$ magnitude for each galaxy. A drop-off magnitude is indicated by the red dash-dotted lines and the $K_{0}$ magnitude of TRGB is indicated by the blue dash-dotted lines. \label{fig:K_distribution}}
\end{figure}

The number of RSGs in each galaxy is listed in Table \ref{tab:parameters of galaxies}, which is a vast increase in comparison with the previous RSG sample mentioned in Section \ref{sec:intro}.  The $K^\mathrm{TRGB}_{0}$ (blue dashed line) is compared with the drop-off magnitude (red dashed line) of the observation for the sample galaxies in Figure \ref{fig:K_distribution}. If the magnitude that is 0.5 mag brighter than the peak magnitude of the $K_{0}$ distribution is considered to be complete, then the sample of RSGs is complete in IC 1613, NGC 147, NGC 185, WLM, SMC and LMC.  For IC 10, Leo A, Pegasus Dwarf and NGC 6822, $K^\mathrm{TRGB}_{0}$ is 0.38, 0.44, 0.43, and 0.47 mag brighter  than the complete magnitude respectively. Correspondingly, the faint RSGs in these four galaxies are slightly incomplete. For Sextans A and Sextans B, as mentioned above, only bright RSGs are detected. In addition, the near-infrared colors $(J-H)_0$ and $(H-K)_0$ of RSGs would be bluer in metal-poor galaxies so that they can become bluer than dwarf stars in the Milky Way galaxy if the metallicity is very low. In such case, RSGs would blend into the area of dwarfs in the near-infrared CCD. The consequence is that some RSGs would be removed by our criteria. Because the color of TRGB depends on metallicity, comparing $(J-K)^\mathrm{TRGB}_{0}$ with that of SMC indicates that  Sextans A, Sextans B, IC10 and WLM should be most influenced by this effect, and their samples of RSGs must be incomplete at the blue end that is the low-mass end. Moreover, none of the sample galaxies is as metal-rich as the Milky Way galaxy and should be more or less affected by this effect. From the metallicity point of view, only M31 and M33 would have a complete sample of RSGs, and the sample of LMC and SMC should also be approximately complete because the foreground dwarfs are removed based on the proper motion independent of metallicity. In conclusion, the samples of RSGs are greatly extended in these galaxies though a high proportion is not yet complete. The full catalogs of RSGs are listed in Table \ref{tab:catalog_dwarf} and \ref{tab:catalog_mcs} with R.A. and Decl. coordinates, magnitudes, magnitude errors, astrometric information, etc.

The pollution rate of the RSGs sample is difficult to estimate by making use of the control fields as done by \citetalias{2021ApJ...907...18R}. The unclear profile and the small size of dwarf galaxies both make the choice of a control field difficult. After removing the foreground stars effectively, the samples of member stars are generally pure. Nevertheless, it is worth noting that AGB stars may shift to the RSG region in the $(J-K)_{0}/K_{0}$ diagram due to the photometric error to bring about contaminants of the RSG sample.

In order to estimate this contamination rate of RSGs by AGBs, the CMD model is constructed as described in Section 5.1 by using the metallicity of LMC, where RSG and AGB stars are marked according to the classification criteria. The error of each RSG and AGB star is obtained from the relation of the photometric error with the apparent magnitude calculated from the model absolute magnitude and the galaxy's distance modulus\footnote{The relation of the photometric error with apparent magnitude takes the form of $\mathrm{error} = A + e^{B(\mathrm{mag}-C)}$. The coefficients are determined from fitting the photometric data of the initial sample. For 2MASS data, $\mathrm{JMag\_Err} = 0.02 + e^{0.97(\mathrm{JMag}-18.76)}$ and $\mathrm{KMag\_Err} = 0.02 + e^{0.91(\mathrm{KMag}-13.31)}$.}. For each source in the CMD model, 1000 simulations are performed with a normal distribution of error whose average is the above calculated value. Finally, the contamination rate of RSGs by AGBs can be calculated based on the simulated sources and the selection criteria.

However, for the UKIRT data of other galaxies, this method may overestimate the contamination rate at large distance modulus. The reason is that the contamination rate increases with magnitude, but at large distance modulus, the faint sources are not detectable. Including these non-detected sources would make the contamination rate higher than the true value. Therefore, for the UKIRT data, a new method is proposed to estimate the contamination rate and completeness. We extract the observational information (i.e., seeing, exposure time, magnitude zero-point, readout noise, etc.) in the Flexible Image Transport System (FITS) header of two typical UKIRT images in the $J$ and $K$ bands. Then the simulated images are generated by the software SkyMaker using the typical values in the FITS header and taking the Paodva CMD as input catalog. The simulated images are used as input to both the CASU imcore\footnote{http://casu.ast.cam.ac.uk/surveys-projects/software-release} program to generate FITS-format catalogs and convert to ASCII-format catalogs using CASU fitsio\_cat\_list program. In this way, we get the simulated CMD with the observational photometric error and the simulated CMD can be used to estimate the contamination rate and completeness at different modulus.

It should be mentioned that in the case of 2MASS and UKIRT the number of RSGs and AGBs has been multiplied by the absolute number of stars occupying the isochrone section per unit mass using the integral of the IMF given in the CMD model\footnote{http://stev.oapd.inaf.it/cmd\_3.5/help.html}. In the case of the 2MASS data, the contamination rate of RSGs by AGBs turns out to be $\sim 10\%$ at $\mu=18.5$. In the case of the UKIRT data, this rate is $\sim 17\%$ at $\mu=24$ and rises to $\sim 30\%$ at $\mu=25$. Definitely, the contamination rate increases towards fainter magnitudes with larger errors. A visual check of the CMD in Figure \ref{fig:fgd_member} would reveal that the samples of RSGs in NGC 147, NGC 185 and NGC 6822 may be significantly contaminated at the faint end, and the present sample is subjected to change with more accurate photometry. On the other hand, the completeness of the RSG sample can be estimated by counting the RSGs that is shifted to the AGB region by the photometric error. For the 2MASS data, the completeness of RSGs turns out to be $\sim 92\%$ at $\mu=18.5$. For the UKIRT data, the completeness becomes $\sim 82\%$ and $\sim 71\%$ at $\mu=24$ and $\mu=25$ respectively. It is worth emphasizing that the completeness of the UKIRT data take into account both the incompleteness brought by the photometric error and the depth of photometry. In addition, there are still some limitations of the simulation that need to be noted: differences in the properties of galaxies make the actual number ratio of RSGs to AGBs vary among galaxies, meanwhile an inaccuracy or spread in metallicity could affect the RSG selection box as well as increase the number of AGB stars straying into it. Finally, the effect of stellar blends is likely to result in bluer (and brighter) sources that could lead to contamination of the RSGs.

\subsubsection{Spatial Distribution} \label{sec:Spatial Distribution}

\begin{figure}
	\centering
    \includegraphics[scale=0.26]{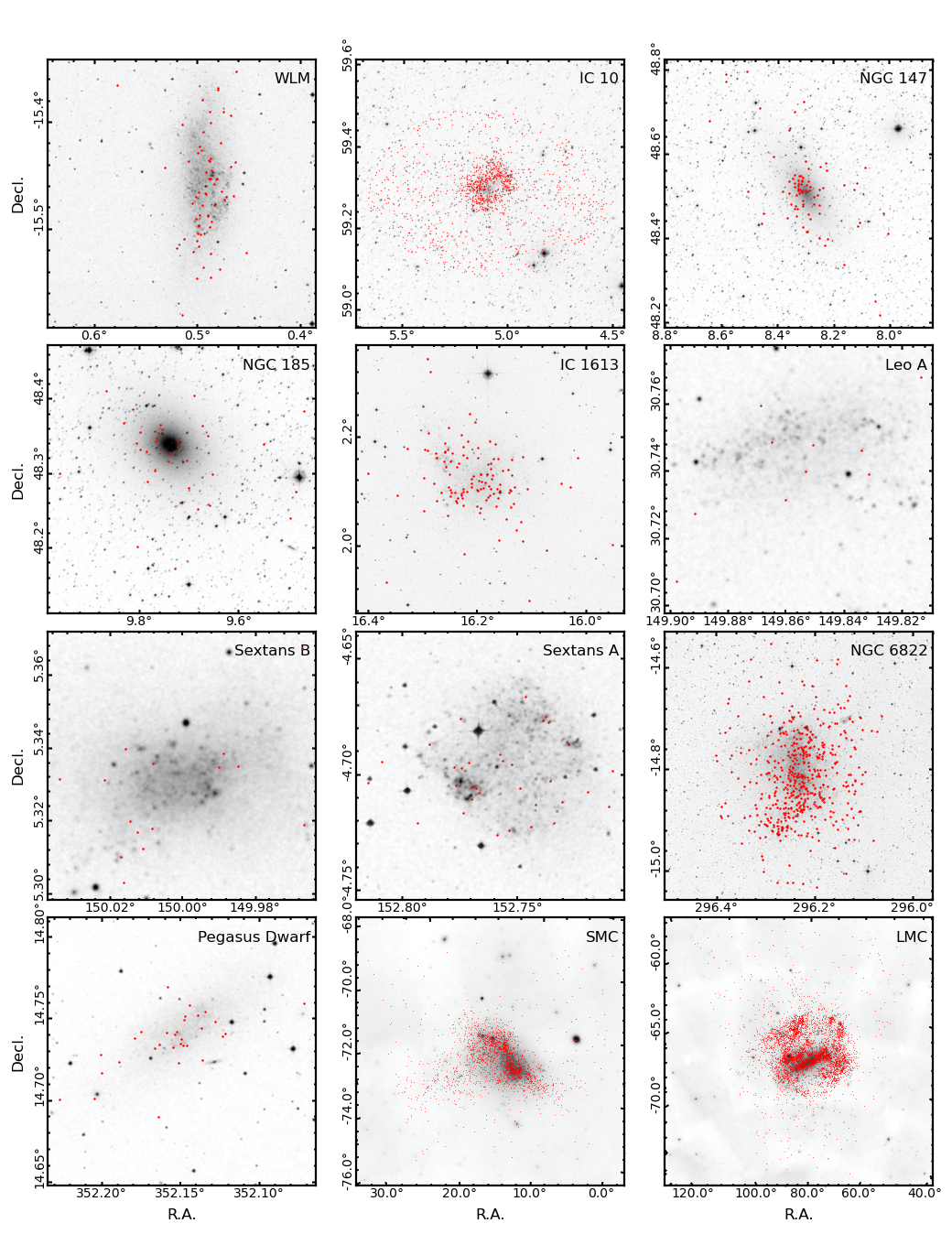}
	\caption{Spatial distribution of the RSGs in each galaxy, with the DSS2 image as the background. \label{fig:spatial_distribution}}
\end{figure}

The spatial distribution of the selected RSGs is shown in Figure \ref{fig:spatial_distribution} with the Digitized Sky Surveys 2 (DSS2) image as the background. It is known that dwarf elliptical galaxies and much more diffuse dwarf spheroidals are gas-free with weak star formation. The distribution of RSGs in such galaxies looks very scattering as demonstrated in NGC 147, NGC 185, Sextans A, Sextans B and Pegasus dwarfs (c.f. Table \ref{tab:basic information} for the galaxy type). By contrast, dwarf irregulars are gas-rich galaxies with evident star-formation activity. The distribution of RSGs presents certain structures in such galaxies, i.e. IC 10, IC 1613, WLM, NGC 6822, SMC and LMC, which should coincide with the star-forming regions. The morphologic study of various stellar populations in IC 1613 \citep{2019MNRAS.483.4751H}, NGC 6822 \citep{2021arXiv210807105K}, SMC and LMC \citep{2019MNRAS.490.1076E} also found that RSGs occupied the bar of the galaxies and they were related to star formation in the last $\sim$200 Myr. \citet{2019MNRAS.490.1076E} noticed that the distribution of RSGs is heavily affected by the presence of Milky Way stars. After removing the foreground stars, the distribution of RSGs can better trace the star-forming region in this work. In fact, the spatial distribution of RSGs in IC 10 is very concentrated in the central region, but shows no more fine structure. This may be due to its active star formation resulting in a large number of RSGs, or due to the inaccurate two-dimensional extinction map. One exception is Leo A, which was classified as a dwarf irregular galaxy, but the identified RSGs exhibit no structure at all. One possible reason is that the sample of RSGs is very small with only ten objects and could be incomplete due to its low metallicity. But it can also be authentically short of massive stars. It is found to be a gas-rich \citep{1996ApJ...462..203Y,2012AJ....144..134H} stellar system dominated by dark matter \citep{2007ApJ...666..231B,2017ApJ...834....9K} of low stellar mass \citep{2007ApJ...659L..17C} and low metallicity \citep{2006ApJ...637..269V,2017ApJ...834....9K}. In the sense of RSGs distribution, Leo A is more like a dwarf ellipsoidal galaxy. In general, the distribution of RSGs agrees with the expectation from the type of galaxy, which supports the correct identification.

\subsubsection{Density of RSGs as a Function of Metallicity} \label{sec:Number of RSGs as a Function of Metallicity}

\citetalias{2021ApJ...907...18R} found that the number of RSGs per stellar mass of a galaxy decreases with metallicity based on the sample of RSGs in SMC, LMC, the Milky Way, M33 and M31. In this work, the samples of RSGs are extended to ten more galaxies of the Local Group and the samples for SMC and LMC are updated, which makes it possible to examine the relation between metallicity and the number of RSGs per stellar mass more systematically and accurately.

In order to characterize the density and the massive star formation rate, the number of RSGs is normalized to stellar mass of the galaxy  with the adopted masses of galaxies in Table \ref{tab:parameters of galaxies}. Different from \citetalias{2021ApJ...907...18R}, here the color index of TRGB ($(J-K)^\mathrm{TRGB}_{0}$) is taken as the indicator of metallicity instead of [M/H] or [O/H]. Previous measurements of metallicity exhibit significant dispersion depending on the method and the type of tracers, and such measurements are even unavailable for some galaxies. On the contrary,  $(J-K)^\mathrm{TRGB}_{0}$ correlates very well the metallicity and is the firsthand parameter with small error by the same method, though $(J-K)^\mathrm{TRGB}_{0}$ is the same for Sextans A and Sextans B based on the fact that  their metallicity are the same within the error range as indicated in \citet{2005AJ....130.1558K}. For M31 and M33, the magnitude and color index of TRGB is determined based on the catalog of member stars from \citetalias{2021ApJ...907...18R} after extinction correction using the Bayestar19 dust map. One point to note is that NGC 147, NGC 185 and Pegasus Dwarf were considered to have smaller metallicity than LMC \citep{2005MNRAS.356..979M,2012AJ....144....4M}, which is different from our result. If the color index of TRGB is determined redder than the true value, it will classify some AGBs as RSGs. For Pegasus Dwarf, there is a potential risk that there are not enough member stars which may make the determination of TRGB difficult. On the other hand, the distance modulus is determined to be 24.97 which agrees with previous results such as 24.82 in \citet{2012AJ....144....4M}, 25.04 in \citet{2010AJ....140.1475P} and 24.96 in \citet{2017ApJ...834...78M}. The determined distance moduli of NGC 147 and NGC 185 in this work also coincide with those of \citet{2012AJ....144....4M}. This implies that the determination of TRGB is reasonable in our work. So further research may be needed to assure the metallicity of Pegasus Dwarf, NGC 147 and NGC 185.

\begin{figure}
	\centering
    \includegraphics[scale=0.5]{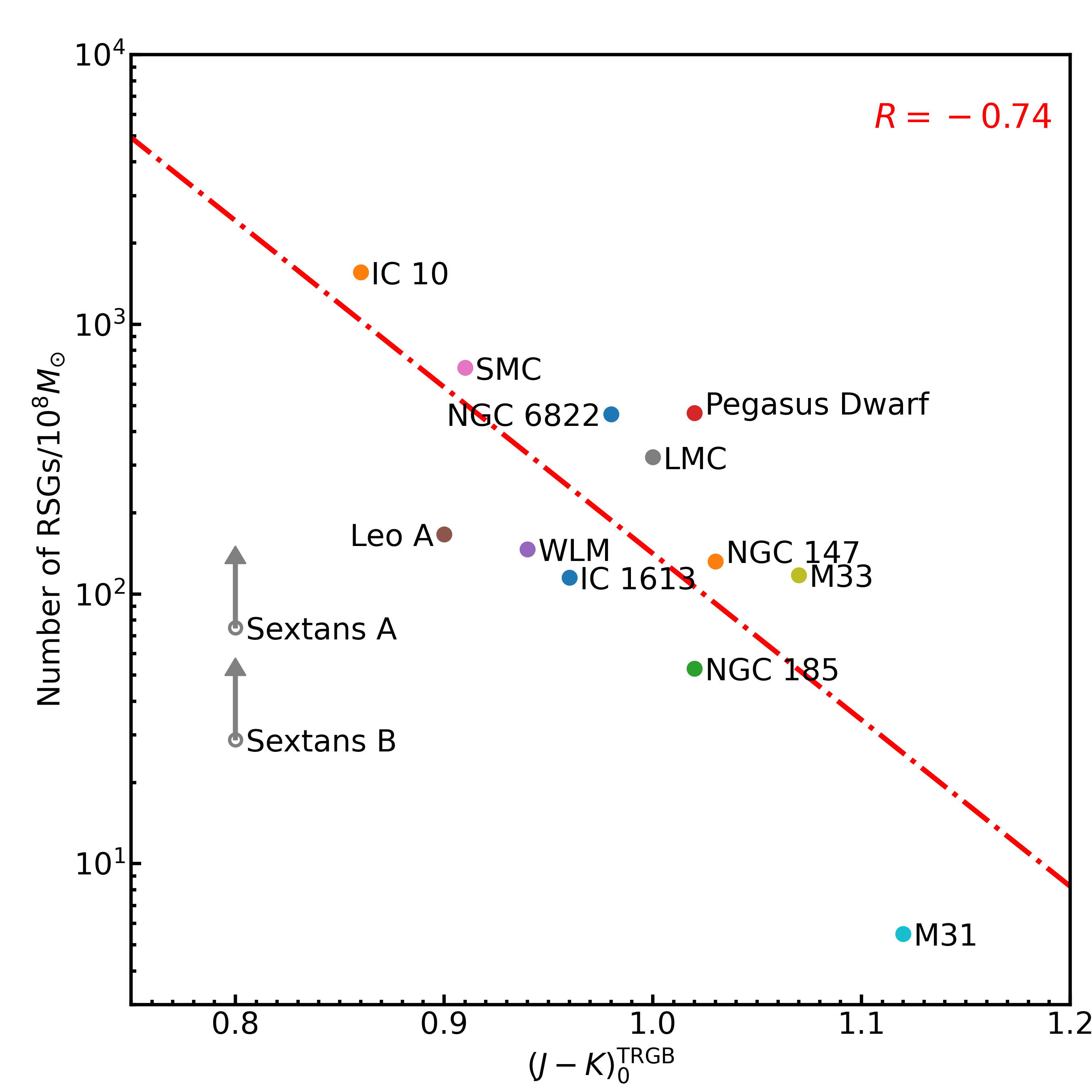}
	\caption{Variation of the number of RSGs per stellar mass with $(J-K)^\mathrm{TRGB}_{0}$ for our studied 14 galaxies. For Sextans A and Sextans B, the samples of RSGs are incomplete by the observational limit and denoted by gray circles and upward arrows. \label{fig:mass_number}}
\end{figure}

With increasing $(J-K)^\mathrm{TRGB}_{0}$ (i.e. metallicity \citealp{2004A&A...424..199B,2018AJ....156..278G}), the RSG density per stellar mass decreases rapidly as shown in Figure \ref{fig:mass_number}. A linear fitting between RSG density per stellar mass and $(J-K)^\mathrm{TRGB}_{0}$ yields the following relation with a high Pearson correlation coefficient $R = -0.74$:

\begin{equation} \label{eq:RSG density}
	\mathrm{log}\left( N_\mathrm{RSGs}/10^8M_{\odot} \right) = -6.17(J-K)^\mathrm{TRGB}_{0} + 8.32.
\end{equation}

This relation is in agreement with \citetalias{2021ApJ...907...18R}. The reason for the decrease of RSGs per stellar mass with metallicity is that the mass loss causes the higher luminosity RSGs to evolve back to blue at higher luminosity \citep{2021arXiv210708304M}. In fact, there is a similar effect for all RSGs. A higher metallicity will bring about a higher mass loss rate \citep{2021ApJ...912..112W} and make RSGs leave RSG phase more quickly. Previous studies have shown that metallicity affects the ratio of blue-to-red supergiants (B/R) and Wolf-Rayet stars to RSGs (WR/RSG). When metallicity increases by 0.9 dex, the B/R ratio and the WR/RSG ratio increase by about 7 times \citep{1980A&A....90L..17M} and 100 times \citep{2002ApJS..141...81M} respectively. The decrease in the number of RSGs per stellar mass is one of the reasons that the WR/RSG ratio increase with metallicity \citep{2021arXiv210708304M} (another is the number of WRs increase with metallicity). The increase of RSG number density can be understood by the metallicity effect on stellar evolution. Nevertheless, the star formation rate and the metallicity effect on initial mass function can play important role in the number of RSGs in a galaxy, which will be investigated in our future work.

\subsection{Comparison with Previous Results} \label{sec:Comparison with Previous Results}

\subsubsection{The Magellanic Clouds} \label{sec:MCs_Comparison}

This work identifies 4,823 and 2,138 RSGs in LMC and SMC respectively, an apparently much larger sample than the 2,974 and 1,239 RSGs by \citet{2021AandA...646A.141Y} and \citet{2020AandA...639A.116Y}.

For LMC (SMC, the following numbers in the brackets refers to SMC), 2,286 (for SMC 921) out of 4,823 (for SMC 2,138) RSGs in our sample can be matched with \citet{2021AandA...646A.141Y} and \citet{2020AandA...639A.116Y} and 2,537 (for SMC 1,217) are new in this work. On the other hand, 688 (for SMC 318) out of 2,974 (for SMC 2,138) RSGs in \citet{2021AandA...646A.141Y} are not included in our sample. These 688 (for SMC 318) RSGs are distributed on both sides of the common sample in the $(J-K)_{0}\ vs.\ K_{0}$ diagram, which can be caused by the slightly different selection of the red and blue boundaries of the RSGs in the two works. In addition, the interstellar extinction is corrected in this work while not in \citet{2021AandA...646A.141Y}. As the reddened RSGs would move in and a few reddened yellow supergiants would move out of the boundaries after correcting interstellar extinction, this would cause the difference in the two samples. The 2,537 (for SMC 1,217) new RSGs mostly fall within the region of \citet{2021AandA...646A.141Y} for RSGs (red boundary, blue boundary, and lower limit are roughly the same). The main reason that they are missed by \citet{2021AandA...646A.141Y} lies in the difference of the initial sample. \citet{2021AandA...646A.141Y} obtained the initial sample by cross-matching the catalog of Spitzer Enhanced Imaging Products (SEIP) including the 2MASS data and the Gaia/DR2 catalog with a 3'' deblending and a 10$\sigma$ detection limit of SEIP data, which leads to losing part of the near-infrared data and finally yields the initial sample of 264,292 (for SMC 74,237) sources, while the initial sample in this work contains 776,831 (for SMC 133,843) sources. Though a larger sky area can account for part of this increase, our initial sample would still be twice that of \citet{2021AandA...646A.141Y} even if the sky area range is restricted to be consistent with previous work, which explains largely the increase of the number of RSGs. Besides, the Gaia/EDR3 provides better measurements to many more stars in the LMC area.

Our sample can be purer than previous samples in that the objects selected by proper motions as before are further purified by the NIR CCD to remove additional foreground stars. Checking the numbers in Table \ref{tab:numbers} find that about 15\% foreground dwarf stars are removed only by the NIR CCD criteria, in other words, about 15\% foreground stars contaminate the sample of member stars (including the RSGs) selected only based on the proper motions for LMC and SMC.


\subsubsection{NGC 6822} \label{sec:NGC6822_Comparison}

In NGC 6822, 465 RSGs are identified, about twice of 234 RSGs by \citet{2021A&A...647A.167Y}. Among them, 135 RSGs are common with \citet{2021A&A...647A.167Y}. About 40 of the 99 sources unique to \citet{2021A&A...647A.167Y} are considered to be foreground dwarfs in the $(J-H)_{0}\ vs.\ (H-K)_{0}$ diagram. Some of them may be true RSGs in NGC 6822 because they may have color indices similar to foreground dwarf stars due to the poor metallicity of NGC 6822. On the other hand, there are an additional 330 RSGs in this work, about 90 of which is increased by the selection of a lower limit by $\sim0.4$ mag in $K_{0}$ for RSGs than \citet{2021A&A...647A.167Y}. Besides, \citet{2021A&A...647A.167Y} required the measurements in the PS1 $rz$ band to discriminate the giant from dwarf stars, which reduces apparently the sample of member stars.

\section{Summary} \label{sec:summary and conclusion}

This work selects the red supergiants in twelve low-mass galaxies of the Local Group by using mainly the near-infrared photometry data in the $JHK$ bands taken by UKIRT/WFCAM from mid-2005 to 2007 and the 2MASS point source catalog as well as the PS1 photometry and Gaia astrometric information. After the interstellar extinction is corrected by various methods, the foreground dwarf stars are removed according to their apparent branches in the  $(J-H)_{0}\ vs.\ (H-K)_{0}$ CCD due to the more significant darkening in the $H$ band of dwarf than giant stars. For the Large and Small Magellanic Clouds, the rejection of foreground stars is mainly based on the reliable and almost complete measurements of the proper motion and parallax by Gaia/EDR3. In the $(J-K)_0/K_0$ diagram of the member stars, RSGs are identified by their larger luminosity and bluer color than TRGB, where the boundaries in this color-magnitude diagram are defined by the most reliable measurements of LMC.

The analysis identified a total of 2,190 RSGs in ten galaxies, and additionally 4,823 and 2,138 RSGs in the Large and Small Magellanic Clouds, respectively. By comparing the photometric completeness depth with the lower brightness limit of RSGs, the sample of RSGs are complete for all the sample galaxies but Sextans A and Sextans B. However, the RSGs in metal-poor galaxies may have near-infrared color bluer than the foreground dwarf stars, this may lead to missing some RSGs in the metal-poor galaxies such as Sextans A, Sextans B, IC10 and Leo A. This problem calls for a new method to remove foreground dwarf stars for metal-poor galaxies.

Based on the sample of the member stars, the magnitude $K_{0}$ and color index $(J-K)_{0}$ of the TRGB are determined by the Poisson Noise weighted star counts method. The color index $(J-K)_{0}$ of the TRGB is an indicator of the metallicity, which aligns the relative metallicity sequence for these sample galaxies. When taking $(J-K)_{0}$ as the metallicity index, it is confirmed that the number of RSGs per stellar mass decreases with increasing metallicity. In addition, $K_{0}$ of the TRGB is used to calculate the distance modulus of the galaxy. While this distance modulus coincides very well within the dispersion of previous results, it is smaller than the highly consistent value of LMC and SMC, which demands improving the calibration of the color-metallicity-magnitude relation of TRGB.

\acknowledgments{We thank Drs. Yang Chen, Xiaoting Fu, Jian Gao, Haibo Yuan, He Zhao and Mingxu Sun for their very helpful discussions, and Dr. Jacco van Loon for very helpful suggestions. We are grateful to Dr. Mike Read for his kind assistance with the UKIRT data.
This work is supported by National Key R\&D Program of China No. 2019YFA0405503, CMS-CSST-2021-A09 and NSFC 12133002. This work has made use of data from UKIRT, 2MASS, PS1, LGGS and Gaia.}

%

\vspace{5mm}
\facilities{}


\software{Astropy \citep{2013A&A...558A..33A},
		  TOPCAT \citep{2005ASPC..347...29T},
		  LMfit \citep{2018zndo...1699739N},
		  dustmaps \citep{2018JOSS....3..695M},
		  SkyMaker \citep{2009MmSAI..80..422B}
          }




\bibliography{paper}{}
\bibliographystyle{aasjournal}


\begin{deluxetable*}{crrrrcccc} \label{tab:basic information}
	\tablecaption{Basic Information of the Twelve Studied Galaxies $^{e}$}
	\tablewidth{0pt}
	\tablehead{
		\colhead{Name} & \colhead{R.A. (J2000)} & \colhead{Decl. (J2000)} & \colhead{$l$} & \colhead{$b$} & \colhead{Type} & \colhead{Diameter} & \colhead{$\mathrm{A}_{V}^{\rm Foregpround}$} & \colhead{d.m. ($\mu$)} \\
		\colhead{} & \colhead{(deg)} & \colhead{(deg)} & \colhead{(deg)} & \colhead{(deg)} & \colhead{} & \colhead{(arcsec)} & \colhead{(mag)} & \colhead{(mag)}
	}
	\startdata
	WLM & 0.4923 & $-$15.4609 & 75.8629 & $-$73.6243 & dIrr & 704.90 & 0.104 & 24.85$^{a}$ \\
	IC 10 & 5.0723 & 59.3038 & 118.9590 & $-$3.3274 & dIrr & 809.40 & 4.299 & 24.50$^{a}$ \\
	NGC 147 & 8.3005 & 48.5088 & 119.8174 & $-$14.2526 & dE & 900.00 & 0.473 & 24.15$^{a}$ \\
	NGC 185 & 9.7415 & 48.3374 & 120.7918 & $-$14.4825 & dE & 929.30 & 0.505 & 23.95$^{a}$ \\
	IC 1613 & 16.1991 & 2.1178 & 129.7377 & $-$60.5773 & dIrr & 1320.00 & 0.068 & 24.39$^{a}$ \\
	Leo A & 149.8603 & 30.7464 & 196.9037 & 52.4225 & dIrr & 330.00 & 0.057 & 24.51$^{a}$ \\
	Sextans B & 150.0004 & 5.3322 & 233.2001 & 43.7838 & dSph & 360.00 & 0.085 & 25.77$^{a}$ \\
	Sextans A & 152.7533 & $-$4.6928 & 246.1482 & 39.8756 & dSph & 353.30 & 0.122 & 25.78$^{a}$ \\
	NGC6822 & 296.2406 & $-$14.8034 & 25.3394 & $-$18.3992 & dIrr & 1117.30 & 0.646 & 23.40$^{b}$ \\
	Pegasus Dwarf & 352.1510 & 14.7429 & 94.7765 & $-$43.5542 & dIrr/dSph & 307.70 & 0.187 & 24.82$^{a}$ \\
	SMC & 13.1866 & $-$72.8286 & 302.7969 & $-$44.2992 & Irr & 19867.90 & 0.101 & 18.95$^{c}$ \\
	LMC & 80.8939 & $-$69.7561 & 280.4653 & $-$32.8883 & SBm/Irr & 41509.80 & 0.206 & 18.49$^{d}$ \\
	\enddata
\tablecomments{$^{a}$\citealt{2012AJ....144....4M} \\
$^{b}$\citealt{2012MNRAS.421.2998F,2014ApJ...794..107R} \\
$^{c}$\citealt{2014ApJ...780...59G,2016ApJ...816...49S} \\
$^{d}$\citealt{2013Natur.495...76P} \\
$^{e}$\href{http://ned.ipac.caltech.edu}{http://ned.ipac.caltech.edu}
}
\end{deluxetable*}

\begin{deluxetable*}{ccccc}[ht] \label{tab:ellipses}
	\tablecaption{Parameters of the Proper Motion Ellipses for SMC and LMC}
	\tablewidth{0pt}
	\tablehead{
		\colhead{} & \colhead{Center} & \colhead{Major Axis} & \colhead{Minor Axis} & \colhead{Position Angle$^{a}$} \\
		\colhead{} & \colhead{(mas/yr, mas/yr)} & \colhead{(mas/yr)} & \colhead{(mas/yr)} & \colhead{(deg)}
	}
	\startdata
	SMC & (0.682, $-$1.251) & 2.086 & 1.328 & $-$4.326 \\
	LMC & (1.814, 0.335) & 4.022 & 2.310 & $-$75.662 \\
	\enddata
\tablecomments{$^{a}$Rotation in degrees anti-clockwise.}
\end{deluxetable*}

\begin{deluxetable*}{ccccccc}[ht] \label{tab:numbers}
	\tablecaption{Number of Stars in the Initial Sample, the Foreground Removed and the Member}
	\tablewidth{0pt}
	\tablehead{
		\colhead{Name} & \colhead{Initial sample/With Gaia$^{a}$/With PS1$^{b}$} & \multicolumn4c{Removed Foreground stars}  & \colhead{Member Stars} \\ \hline
\colhead{} & \colhead{} & \colhead{by UKIRT/2MASS} & \colhead{by PS1} & \colhead{by Gaia} & \colhead{Total} & \colhead{}
	}
	\startdata
	WLM & 1,651/275/598 & 735 & 156 & 31 & 761 & 890 \\
	IC 10 & 41,063/11,429/11,043 & 25,614 & 10,875 & 1,472 & 25,727 & 15,336 \\
    NGC 147 & 20,252/3,374/4,591 & 9,463 & 3,248 & 629 & 9,685 & 10,567 \\
    NGC 185 & 16,752/2,984/4,166 & 8,446 & 3,026 & 528 & 8,737 & 8,015 \\
	IC 1613 & 4,876/1,280/3,399 & 2,385 & 761 & 142 & 2,513 & 2,363 \\
	Leo A & 199/70/141 & 101 & 51 & 5 & 113 & 86 \\
	Sextans B & 272/86/95 & 127 & 58 & 13 & 131 & 141 \\
    Sextans A & 403/156/216 & 189 & 105 & 23 & 209 & 194 \\
	NGC 6822 & 12,308/7,017/8,091 & 7,789 & 4,239 & 669 & 8,224 & 4,084 \\
    Pegasus Dwarf & 693/97/250 & 291 & 79 & 20 & 295 & 398 \\
    SMC & 133,843/132,581/-- & 38,664 & -- & 57,532 & 67,270 & 66,573 \\
    LMC & 776,831/768,192/-- & 283,760 & -- & 314,656 & 374,531 & 402,300 \\
	\enddata
\tablecomments{$^{a}$Number of sources with Gaia/EDR3 astrometric information. \\
				$^{b}$Number of sources with PS1 ``good measurement''.}
\end{deluxetable*}

\begin{deluxetable*}{ccccccc}[ht] \label{tab:parameters of galaxies}
	\tablecaption{Number of Red Supergiants, Position of TRGB and Distance Modulus of the Galaxies}
	\tablewidth{0pt}
	\tablehead{
		\colhead{Name} & \colhead{Number of RSGs} & \colhead{Stellar Mass} & \colhead{$K^\mathrm{TRGB}_{0}$} & \colhead{$(J-K)^\mathrm{TRGB}_{0}$} & \colhead{$\mu$}\\
		\colhead{} & \colhead{} & \colhead{($10^{8}M_{\odot}$)} & \colhead{(mag)} & \colhead{(mag)} & \colhead{(mag)}
	}
	\startdata
	WLM & 63 & $0.43^{b}$ & 18.68 & 0.94 & 24.77 \\
	IC 10 & 1340 & $0.86^{b}$ & 18.14 & 0.86 & 24.01 \\
	NGC 147 & 82 & $0.62^{b}$ & 17.87 & 1.03 & 24.21 \\
	NGC 185 & 36 & $0.68^{b}$ & 17.62 & 1.02 & 23.94 \\
	IC 1613 & 115 & $1^{b}$ & 18.10 & 0.96 & 24.25 \\
	Leo A & 10 & $0.06^{b}$ & 18.56 & 0.90 & 24.54 \\
	Sextans B & 15 & $0.52^{b}$ & 20.21 & 0.80 & 26.15 \\
	Sextans A & 33 & $0.44^{b}$ & 20.01 & 0.80 & 25.95 \\
	NGC 6822 & 465 & $1^{b}$ & 17.38 & 0.98 & 23.58 \\
	Pegasus Dwarf & 31 & $0.066^{b}$ & 18.65 & 1.02 & 24.97 \\
	SMC & 2138 & $3.1^{c}$ & 12.71 & 0.91 & 18.72 \\
	LMC & 4823 & $15^{c}$ & 12.00 & 1.00 & 18.26 \\
	M31 & $5498^{a}$ & $1000^{d}$ & 17.66$^{e}$ & 1.12$^{e}$ & 24.25\\
	M33 & $3055^{a}$ & $26^{d}$ & 18.17$^{e}$ & 1.07$^{e}$ & 24.62
	\enddata
	\tablecomments{$^{a}$\citetalias{2021ApJ...907...18R} \\
				   $^{b}$\citet{2012AJ....144....4M} \\
				   $^{c}$\citet{2015arXiv151103346B} \\
				   $^{d}$\citet{2016MNRAS.457..844F} \\
				   $^{e}$The magnitude and color index of TRGB are derived based on the member stars in M31 and M33 from \citetalias{2021ApJ...907...18R} after correcting the extinction with the SFD reddening map.
				   }
\end{deluxetable*}

\begin{deluxetable*}{crrccccccccc}[ht] \label{tab:catalog_dwarf}
	\tablecaption{Catalog of Red Supergiants in the Ten Dwarf Galaxies}
	\tablewidth{0pt}
	\tablehead{
		\colhead{Galaxy} & \colhead{R.A.} & \colhead{Decl.} & \colhead{JMag} & \colhead{JMag\_Err} & \colhead{HMag} & \colhead{HMag\_Err} & \colhead{KMag} & \colhead{KMag\_Err} & \colhead{N\_Flag} & \colhead{...} & \colhead{$E(B-V)$}\\
		\colhead{} & \colhead{(deg)} & \colhead{(deg)} & \colhead{(mag)} & \colhead{(mag)} & \colhead{(mag)} & \colhead{(mag)} & \colhead{(mag)} & \colhead{(mag)} & \colhead{} & \colhead{...} & \colhead{(mag)}
	}
	\startdata
	WLM & 0.48253 & $-$15.477478 & 16.845 & 0.01 & 16.152 & 0.01 & 16.055 & 0.011 & 3 & ... & 0.04 \\
	WLM & 0.479939 & $-$15.370249 & 19.198 & 0.069 & 18.547 & 0.06 & 18.501 & 0.106 & 3 & ... & 0.03 \\
	WLM & 0.498303 & $-$15.465628 & 18.031 & 0.023 & 17.339 & 0.027 & 17.156 & 0.03 & 3 & ... & 0.04 \\
	WLM & 0.489016 & $-$15.435648 & 18.848 & 0.055 & 18.164 & 0.061 & 18.035 & 0.07 & 3 & ... & 0.04 \\
	WLM & 0.519332 & $-$15.517984 & 18.996 & 0.055 & 18.277 & 0.063 & 18.237 & 0.079 & 3 & ... & 0.03
	\enddata
	\tablecomments{(This table is available in its entirety in machine-readable form.)
				   }
\end{deluxetable*}

\begin{deluxetable*}{crrcccccccc}[ht] \label{tab:catalog_mcs}
	\tablecaption{Catalog of Red Supergiants in the Magellanic Clouds}
	\tablewidth{0pt}
	\tablehead{
		\colhead{Galaxy} & \colhead{R.A.} & \colhead{Decl.} & \colhead{JMag$^{a}$} & \colhead{JMag\_Err} & \colhead{HMag$^{a}$} & \colhead{HMag\_Err} & \colhead{KMag$^{a}$} & \colhead{KMag\_Err} & \colhead{...$^{b}$} & \colhead{$E(B-V)$}\\
		\colhead{} & \colhead{(deg)} & \colhead{(deg)} & \colhead{(mag)} & \colhead{(mag)} & \colhead{(mag)} & \colhead{(mag)} & \colhead{(mag)} & \colhead{(mag)} & \colhead{...} & \colhead{(mag)}
	}
	\startdata
	SMC & 10.892176 & $-$73.223694 & 12.476 & 0.021 & 11.748 & 0.025 & 11.613 & 0.023 & ... & 0.06 \\
	SMC & 12.16465 & $-$73.531876 & 11.317 & 0.019 & 10.521 & 0.025 & 10.329 & 0.021 & ... & 0.04 \\
	SMC & 14.998343 & $-$71.675682 & 9.698 & 0.019 & 8.977 & 0.02 & 8.739 & 0.019 & ... & 0.05 \\
	SMC & 12.197104 & $-$73.529991 & 13.288 & 0.023 & 12.599 & 0.029 & 12.485 & 0.029 & ... & 0.04 \\
	SMC & 8.494317 & $-$72.84771 & 10.759 & 0.021 & 9.986 & 0.027 & 9.775 & 0.023 & ... & 0.04
	\enddata
	\tablecomments{$^{a}$2MASS photometric system. \\
				   $^{b}$$JHK$ magnitudes after extinction correction are in UKIRT photometric system. (This table is available in its entirety in machine-readable form.)
				   }
\end{deluxetable*}

\end{CJK*}
\end{document}